\documentclass[a4paper]{JHEP3}

\usepackage{graphicx}
\usepackage{amsmath}
\usepackage{cite}
\usepackage{axodraw}

\usepackage{multirow}
\usepackage{array}

\def\eqref#1{(\ref{#1})}
\def\text{\rm }

\def\mathematica{{{\sc mathematica}}}
\def\sam{{{\sc S@M}}}
\def\form{{{\sc form}}}

\def\C++{{{\sc c++}}}

\newcommand{\beq}{\begin{equation}}
\newcommand{\eeq}{\end{equation}}
\newcommand{\bqa}{\begin{eqnarray}}
\newcommand{\eqa}{\end{eqnarray}}
\newcommand{\bite}{\begin{itemize}}
\newcommand{\eite}{\end{itemize}}

\def\db#1{ D_{#1}}

\def\zzeta{{\mathbf{z}}}

\def\emm{{m}}
\def\idea{{\Gamma}}
\def\ci#1#2{c_{#1,\, #2}}
\def\hci#1#2{\hat c_{#1,\, #2}}
\def\ni#1#2{n_{#1,\, #2}}
\def\hni#1#2{\hat n_{#1,\, #2}}
\def\bi#1#2{b_{#1,\, #2}}
\def\hbi#1#2{\hat b_{#1,\, #2}}

\def\ra{\rangle}

\def\spa#1.#2{\langle#1\,#2\rangle}
\def\spb#1.#2{[#1\,#2]}

\def\spab#1.#2.#3{\langle\mskip-1mu{#1}
                  | #2 | {#3}]}

\def\spba#1.#2.#3{[\mskip-1mu{#1}
                  | #2 | {#3}\rangle}

\def\spbb#1.#2.#3.#4{[\mskip-1mu{#1}
                     | {#2} \ {#3} | {#4}]}

\def\spaa#1.#2.#3.#4{\langle\mskip-1mu{#1}
                     | {#2} \ {#3} | {#4}\rangle}

\newcommand{\bea}{\begin{eqnarray}}

\newcommand{\eea}{\end{eqnarray}}

\newcommand{\bean}{\begin{eqnarray*}}

\newcommand{\eean}{\end{eqnarray*}}
           
\newcommand{\nn}{\nonumber \\}

\newtheorem{prop}{Proposition}[section]

\newtheorem{theo}{Theorem}[section]

\title{Integrand Reduction for Two-Loop Scattering Amplitudes through 
Multivariate Polynomial Division}

\author{Pierpaolo Mastrolia \\
Max-Planck Institut f\"ur Physik, F\"ohringer Ring, 6, D-80805 M\"unchen, Germany

\vspace{1mm}
Dipartimento di Fisica e Astronomia, Universit\`a di Padova, and INFN                                    
Sezione di Padova, via Marzolo 8, 35131 Padova, Italy

\vspace{1mm}
E-mail: \email{ppaolo@mppmu.mpg.de}
}    

\author{Edoardo Mirabella \\
Max-Planck Institut f\"ur Physik, F\"ohringer Ring, 6, D-80805 M\"unchen, Germany

\vspace{1mm}
E-mail: \email{mirabell@mppmu.mpg.de}
}    

\author{Giovanni Ossola \\ 
Physics Department, New York City College of Technology, 
City University of New York,  300 Jay Street, Brooklyn, NY 11201, USA

\vspace{1mm}
Kavli Institute for Theoretical Physics, University of California, Kohn Hall, Santa Barbara, CA 93106, USA

\vspace{1mm}
E-mail: \email{GOssola@citytech.cuny.edu}
}

\author{Tiziano Peraro \\
Max-Planck Insitut f\"ur Physik, F\"ohringer Ring, 6, D-80805 M\"unchen, Germany

\vspace{1mm}
E-mail: \email{peraro@mppmu.mpg.de}
}

\abstract{
We describe the application of a novel approach for the reduction of scattering
amplitudes, based on multivariate polynomial division, which we have
recently presented.
This technique yields the complete integrand decomposition for
arbitrary amplitudes,
regardless of the number of loops.
It allows for the determination of the residue at
any multiparticle cut,
whose knowledge is a mandatory prerequisite for applying
the integrand-reduction procedure.
By using the division modulo Gr\"obner basis,
we can derive a simple integrand recurrence relation that
generates the multiparticle pole decomposition for integrands of  arbitrary
multiloop amplitudes.
We apply the new reduction algorithm to the two-loop planar and
nonplanar diagrams contributing to the  five-point scattering
amplitudes in $\mathcal{N}= 4$ SYM and $\mathcal{N} = 8$ SUGRA in four dimensions,
whose numerator functions contain up to rank-two terms in the
integration momenta.
We determine all polynomial residues parametrizing
the cuts of the corresponding topologies and subtopologies.
We obtain the integral basis for the decomposition of each diagram from  the polynomial form
of the residues. Our approach is well suited for a seminumerical implementation,
and its general mathematical properties provide an
effective algorithm
for the generalization of the integrand-reduction method to all orders in
perturbation theory.
}

\begin{document}

\section{Introduction}
\label{sec:Introduction}

The unitarity of the $S$ matrix encodes the most profound property of a quantum system, namely 
the probability conservation.
The optical theorem, that relates the difference between the
transition amplitude and its complex conjugate  to their product,
is the direct consequence of unitarity.  Hence, at a given order in perturbation theory, 
it connects the discontinuity of the amplitude across a given branch cut 
to the sum of all the Feynman diagrams sharing that specific cut, 
which factorize into two lower-order amplitudes.  
By elaborating on the role of the optical theorem, and introducing the concept of generalised
cuts ~\cite{Bern:1994zx,Britto:2004nc,Cachazo:2004kj,Britto:2004ap}, unitarity has been inspiring a novel organization of the perturbative calculus,
where Feynman diagrams are grouped according to their multiparticle factorization channels.

Scattering amplitudes in quantum field theories are analytic functions of the momenta of the
interacting particles; hence they are determined by their singularities. 
The singularity structure is retrieved when virtual particles go on shell, under the effect 
of complex deformations of the kinematic, as needed for solving multiple on-shell 
conditions simultaneously.

The  investigation of the mathematical properties of the residues at the 
singularities 
led to the discovery of
new relations involving scattering amplitudes, such as 
the BCFW  recurrence relation \cite{Britto:2004ap},
its link to the leading singularity of one-loop amplitudes \cite{Britto:2004nc},
and the  OPP integrand-decomposition formula~\cite{Ossola:2006us}. 

Automating the evaluation of one-loop multiparticle amplitudes, for
an accurate description of scattering processes that were
considered prohibitive, has become feasible.
Motivated by the challenging experimental program of the LHC, where the ubiquity of QCD
manifests itself through the production of multijet events, 
several codes have been developed with the goal
of reaching the next-to-leading order level of accuracy for the cross sections 
\cite{Berger:2008sj,Giele:2008bc,Badger:2010nx,Bevilacqua:2011xh,
Hirschi:2011pa,  Cullen:2011ac,Agrawal:2011tm,Cascioli:2011va,Badger:2012pg}.

On the more mathematical side, it became clear that within the
on-shell and unitarity-based methods,
the theory of multivariate complex functions could play an important role 
in order to compute the generalized cuts efficiently.
The holomorphic anomaly~\cite{Cachazo:2004dr,Britto:2004nj} and the spinor 
integration~\cite{Britto:2005ha,Britto:2006sj}, as well as, 
Cauchy's residue theorem \cite{Britto:2004ap,Britto:2004nc},
Laurent series expansion \cite{Forde:2007mi,Badger:2008cm,ArkaniHamed:2008gz},
Stokes' Theorem \cite{Mastrolia:2009dr,Britto:2010um}, and Global residue theorem
\cite{ArkaniHamed:2009dn} have been employed
for carrying out the integration of the phase-space
integrals, left over after applying the on-shell cut conditions to the
loop integrals.

Progress on the unitarity-based methods 
and the vivid research activity spun off 
has been recently reviewed in~\cite{Ellis:2011cr} and~\cite{Alday:2008yw,Britto:2010xq,Henn:2011xk,Bern:2011qt,Carrasco:2011hw,Dixon:2011xs,Ita:2011hi}.

At two loops, generalized unitarity techniques have been introduced for supersymmetric amplitudes~\cite{Bern:1997nh} and later  for  
QCD amplitudes~\cite{Bern:2000dn}. The multiple cuts of  two-loop amplitudes were proposed to extend the simplicity of 
the one-loop quadruple cut~\cite{Britto:2004nc}  to the leading singularity techniques~\cite{Buchbinder:2005wp,Cachazo:2008vp} 
and to the maximal-cuts method~\cite{Bern:2007ct}.  The maximal-unitarity approach developed by Kosower, Larsen, Caron-Huot,
and Johansson~\cite{Kosower:2011ty,Larsen:2012sx,CaronHuot:2012ab,Johansson:2012zv} has refined this technique by a 
 systematic application of the global residue theorem. \\

The singularity structure of multiloop scattering amplitudes
can be also exposed in the integrand,
before integrating over the loop momenta.
The integrand-reduction methods
use the singularity structure of the integrands  to decompose the (integrated) amplitudes in terms of Master Integrals (MIs). 
The multiparticle pole expansion of the integrand 
is equivalent to the decomposition of the numerator in terms of 
products of denominators, multiplied by polynomials. These latter
correspond to the residues at the multiple-cuts. 
In general, the coefficients of the MIs are a subset of the
coefficients appearing in the polynomial residues. 
Therefore the complete determination of the residues leads to the
complete decomposition of the amplitudes in terms of MI's.  The final result
is then obtained by evaluating the latter.
%
%
%

The parametric form of the polynomial residues  is process independent
and it can be determined  {\it a priori}, from the topology of the corresponding
on-shell diagram, namely from the graph identified by the denominators that 
go simultaneously on shell. The actual value of the  coefficients is clearly  process dependent, and 
its determination is indeed the goal of the integrand reduction. Integrand-reduction methods 
determine the (unknown) coefficients by {\it polynomial fitting}, through the 
evaluation of the (known) integrand at values of the
loop momenta fulfilling the cut conditions. 
The integrands contributing to the amplitude, which have to be
evaluated in correspondence to the solutions of the on-shell
conditions, are the only input required. They can be provided either
as a product of tree-level amplitudes, like in unitarity-based
approaches, or as a combination of Feynman diagrams, retaining the
full loop-momentum dependence. In the former case the
 on-shell diagram represents a cut of the amplitude while 
 in the latter case it is simply the cut of an
integral where the on-shell conditions are applied to its numerator.
The integrand-reduction methods have been originally developed at one loop \cite{Ossola:2006us}. 
Extensions  beyond 
one loop were proposed in~\cite{Mastrolia:2011pr,Badger:2012dp}.
A key point of the higher-loop extension is the proper parametrization
of the residues of the multiparticle poles.
Each residue is a multivariate polynomial in the 
{\it irreducible scalar products} (ISPs) among the loop momenta and 
either external momenta or polarization vectors constructed out of
them. ISPs cannot be expressed in terms of denominators, thus 
any monomial formed by ISPs is the numerator of an integral which 
may be a MI  appearing in the final result.

Both the numerator and the denominators of any integrand are
multivariate polynomials in the components of the loop variables.
As recently shown in~\cite{Zhang:2012ce,Mastrolia:2012an},  
the decomposition of the integrand can be
obtained using basic principles of algebraic geometry, by performing the
{\it multivariate polynomial division} between the numerator and the Gr\"obner
basis generated by (a subset of) the denominators.
Moreover,  the multivariate polynomial divisions  give a systematic classification 
of the polynomial structures of the residues,  leading to both the identification 
of the MIs and the determination  of their coefficients. 

In~\cite{Mastrolia:2011pr} it was observed that the set of independent integrals which emerge 
from the integrand-reduction algorithms is not minimal.
Integration-by-parts~\cite{Tkachov:1981wb}, Lorentz-invariance~\cite{Gehrmann:2000xj}, 
and  Gram-determinant~\cite{Gluza:2010ws} identities may constitute additional, independent relations 
which can further reduce the number of MI's that have to be actually evaluated, after the reduction 
stage. 
Badger, Frellesvig and Zhang have explicitly shown that the number of independent ten-denominator
integrals identified through the integrand decomposition of the three-loop four-point 
ladder box diagram  is significantly reduced by using integration-by-parts identities~\cite{Badger:2012dv}.
In the case of 
one- and two-loop amplitudes, an alternative technique for 
counting the numbers of tensor structures and of the independent coefficients
has been presented by Kleiss, Malamos, Papadopoulos and Verheyen in~\cite{Kleiss:2012yv}. 
During the completion of this work, 
Feng and Huang \cite{Feng:2012bm} have shown that, 
by using multivariate polynomial division \cite{Zhang:2012ce,Mastrolia:2012an}, 
a systematic classification of a four-dimensional integral basis for two-loop integrands is doable.

In \cite{Mastrolia:2012an}, we have set the mathematical framework for the multiloop integrand-reduction 
algorithm. We have shown that  the residues are
uniquely determined by the denominators involved in the corresponding multiple cut.  
We have derived a simple {\it integrand recurrence relation} 
generating the multiparticle pole decomposition. 
The algorithm is valid  for arbitrary amplitudes, irrespective of  
the number of loops, the particle content (massless or massive), 
and of the diagram topology (planar or nonplanar). Interestingly, 
at one loop our algorithm allows for a simple derivation of the  OPP reduction formula~\cite{Ossola:2006us}. 
The spurious terms, when present, naturally 
arise from the structure of the denominators entering the generalized cuts.

In the same work \cite{Mastrolia:2012an}, 
we gave the proof of the {\it maximum-cut theorem}. The theorem deals with cuts 
where the number of on-shell conditions is equal to the number of integration variables
and therefore the loop momenta are completely localized.
The theorem ensures that the number of independent
solutions of the maximum cut is equal to the
number of coefficients parametrizing the corresponding residue.
The maximum-cut theorem generalizes at any loop the simplicity of  
the one-loop quadruple cut \cite{Britto:2004nc,Ossola:2006us},
where the two coefficients parametrizing the residue
are determined by the two solutions of the cut. \\

In this paper, we apply our algorithm  to the two-loop five-point planar and 
nonplanar diagrams contributing to  amplitudes in ${\cal N}=4$ super Yang-Mills (SYM) and ${\cal N}=8$ Supergravity
(SUGRA) in four dimensions~\cite{Bern:2006ew,Carrasco:2011mn}. We use  the numerator functions  computed in~\cite{Carrasco:2011mn}, 
which  contain  up to rank-two terms in each integration momenta. 
In particular, we derive the generic polynomial residues which are required by the reduction
procedure. Later, we show that the integrand reduction can be performed both 
seminumerically, by polynomial fitting, and analytically.  
The latter computation has been performed generalizing the method of 
integrand reduction through Laurent expansion~\cite{Mastrolia:2012bu}, 
which has been recently introduced to improve the integrand reduction of 
one-loop amplitudes.

All the numerical and analytic computations presented in this paper
have been performed using \C++, \form~\cite{Vermaseren:2000nd} and
the \mathematica~package \sam~\cite{Maitre:2007jq}.

\section{Integrand reduction}
\label{sec:ReductionG}
In this Section we describe the general strategy for the reduction of scattering amplitudes at 
the integrand level, following~\cite{Mastrolia:2011pr,Mastrolia:2012an}.
In dimensional regularization, an $\ell$-loop amplitude can be written as a linear combination 
of  $n$-denominator integrals of the form
\bea
 {\cal A}_n & = & 
\int d^{d}\bar q_1 \ldots  \int d^{d}\bar q_\ell  \quad
   \mathcal{I}_{i_1\cdots i_n}  ( \bar q_1, \ldots ,  \bar q_\ell   ) \nn 
    &\equiv  &  \int d^{d}\bar q_1 \ldots  \int d^{d}\bar q_\ell  \  
\frac{ {\cal N}_{i_1\cdots i_n} ( \bar q_1, \ldots ,  \bar q_\ell   )}{\db{i_1}(\bar q_1, \ldots , \bar q_\ell ) \cdots \db{i_n}( \bar q_1, \ldots  , \bar q_\ell )}\, , \nn
\db{i} &=& \left ( \sum_a \alpha_{i, a} \bar q_a +  p_i \right )^2-m_i^2
\label{def:An}
\eea 
where $q_1, \ldots , q_\ell$ are integration momenta and  $\alpha_{i,a} \in \{0, \pm 1\}$.
Objects living in  $d = 4 - 2\epsilon$ are denoted by a bar.  We  use the notation
$\bar q_a^\mu = q_a^\mu + \vec \lambda_{q_a}$,
where $q_a^\mu$ is the  four-dimensional part of $\bar q_a$, while $\vec \lambda_{q_a}$ 
is its $(-2\epsilon)$-dimensional part~\cite{Bern:2002tk}.
In the following we will limit ourselves to the four-dimensional case.
Extensions to higher-dimensional cases, according to the chosen
dimensional regularization scheme, can be treated analogously.

The integrand-reduction methods~\cite{Ossola:2006us,Ossola:2007bb,Ossola:2007ax,Ossola:2008xq,Ellis:2007br,Giele:2008ve,Ellis:2008ir,Mastrolia:2008jb,Mastrolia:2010nb,Mastrolia:2012bu,Mastrolia:2011pr,Badger:2012dp} trade the decomposition of the loop integrals in terms of MIs with the algebraic problem of building a general relation, at the integrand level, 
for the numerator functions of each integral contributing to  the amplitude. In this paper we use the method introduced in~ \cite{Mastrolia:2012an}.
The algorithm relies solely on general properties of the loop integrand, i.e.  on
the maximum power of the loop momenta present in the numerator, and on
 the quadratic form of Feynman propagators.  
 The residue of  each multiparticle pole
is determined by the on-shell conditions corresponding to the simultaneous vanishing of the
denominators it is sitting on.   In particular, we obtain 
the multipole  decomposition of the  integrand of  Eq.~(\ref{def:An}) using 
an \emph{integrand recurrence relation}  based on   multivariate polynomial division together with 
a criterion for the reducibility  of the integrands.
 In the following subsections we will briefly review the two ingredients of the  method.

\subsection{Integrand recurrence relation}
The four-dimensional version of the integrand of Eq.~(\ref{def:An}) is
 \beq
 \mathcal{I}_{i_1\cdots i_n} \equiv 
\frac{ {\cal N}_{i_1\cdots i_n} ( q_1, \ldots , q_\ell   )}{\db{i_1}(q_1, \ldots ,q_\ell ) \cdots \db{i_n}(q_1, \ldots  ,q_\ell )} \; .
 \label{Eq:IntegrandM}
\eeq
The numerator ${\cal N}_{i_1\cdots i_n}$ and any of the denominators $\db{i}$ 
are polynomial in  the components of the loop momenta, 
 say $\zzeta \equiv (z_1, \ldots z_{4\ell})$, i.e.
 \beq
 \mathcal{I}_{i_1\cdots i_n}  = \frac{{\cal N}_{i_1\cdots i_n}(\zzeta)}{D_{i_1}(\zzeta) \cdots D_{i_n}(\zzeta)} \; . 
 \label{Eq:Igen}
 \eeq
We construct the  ideal generated by the $n$ denominators 
 \bea
{\cal J}_{i_1 \cdots  i_n} &=&  \langle D_{i_1}, \cdots , D_{i_n} \rangle  
\equiv \left \{\sum_{\kappa=1}^n  h_{\kappa}(\zzeta)  D_{i_\kappa}(\zzeta) :  h_\kappa(\zzeta) \in P[\zzeta] \right \} ,
\nonumber 
\eea
 where $P[\zzeta]$ is the set of polynomials in $\zzeta$.
 The common zeros of the elements of 
${\cal J}_{i_1 \cdots  i_n} $ are exactly the common zeros of the denominators. 
We chose a monomial order and  we construct a Gr\"obner basis  generating the ideal ${\cal J}_{i_1 \cdots i_n}$
\beq
\mathcal{G}_{i_1 \cdots  i_n}=\{g_{1}(\zzeta), \ldots  , g_{\emm}(\zzeta) \}  \;  . 
\eeq
The $n$-ple cut conditions $\db{i_1} = \ldots =\db{i_n} = 0$
are equivalent to $g_1 = \ldots = g_\emm = 0$. 
The multivariate division of ${\cal N}_{i_1\cdots i_n}$ modulo 
$\mathcal{G}_{i_1\cdots i_n}$ leads to 
\beq
{\cal N}_{i_1\cdots i_n}(\zzeta) = \idea_{i_1 \cdots  i_n}  + \Delta_{i_1\cdots i_n}(\zzeta)  \; ,
\label{Eq:DecGenI}
\eeq
where $\idea_{i_1 \cdots  i_n} = \sum_{i=1}^\emm  \mathcal{Q}_{i}(\zzeta)
g_i(\zzeta) $ 
is a compact notation for the sum of the products of the quotients $\mathcal{Q}_{i}$ 
and the divisors $g_i$.
The polynomial  $\Delta_{i_1 \cdots i_n}$ is the remainder of the
division. Since ${\cal G}_{i_1 \cdots  i_n}$ is a Gr\"obner basis, 
the remainder is uniquely determined once the monomial order is fixed.
The term $\idea_{i_1 \cdots  i_n}$ belongs to the ideal
${\cal J}_{i_1 \cdots i_n}$,  thus it can be expressed 
in terms of denominators, as
\beq
\idea_{i_1 \cdots  i_n}=  \sum_{\kappa=1}^{n}   
{\cal N}_{i_1\cdots i_{\kappa -1}i_{\kappa+1}\cdots i_n}(\zzeta) \db{i_\kappa}(\zzeta) \, .
\label{Eq:fromGtoD}
\eeq  
The explicit form of 
${\cal N}_{i_1\cdots i_{\kappa -1}i_{\kappa+1}\cdots  i_n}$ can be
found by expressing the elements of the Gr\"obner basis 
in terms of the denominators. 
Using Eqs.~(\ref{Eq:DecGenI}) and~(\ref{Eq:fromGtoD}), we cast the numerator  in the suggestive form
\bea
{\cal N}_{i_1\cdots i_n}(\zzeta) =
\sum_{\kappa=1}^{n}   
{\cal N}_{i_1\cdots i_{\kappa -1}i_{\kappa+1}\cdots i_n}(\zzeta) \db{i_\kappa}(\zzeta) + \Delta_{i_1\cdots i_n}(\zzeta) \ .
\label{Eq:Recursive}
\eea
Plugging Eq.~(\ref{Eq:Recursive}) in Eq.~(\ref{Eq:Igen}), we 
get a nonhomogeneous recurrence  relation for the 
$n$-denominator integrand,
\beq
\mathcal{I}_{i_1\cdots i_n}  =  
 \sum_{\kappa=1}^{n}   \mathcal{I}_{i_1\cdots i_{\kappa -1} i_{\kappa+1} i_n}
+ \frac{\Delta_{i_1\cdots i_n}}{\db{i_1} \cdots  \db{i_n}}  .
\label{Eq:DecGen}
\eeq
According to Eq.~(\ref{Eq:DecGen}), 
$\mathcal{I}_{i_1\cdots i_n}$ is expressed in terms of
$(n-1)$-denominator integrands, 
\bea
\mathcal{I}_{i_1\cdots i_{\kappa -1}  i_{\kappa+1} i_n} = 
\frac{ \mathcal{N}_{i_1\cdots i_{\kappa -1}  i_{\kappa+1} i_n}  }{\db{1}  \cdots \db{i_{\kappa -1}}   \db{i_{\kappa +1}} \cdots \db{i_n}       } \, .
\eea
The nonhomogeneous term contains the remainder of the division~(\ref{Eq:DecGenI}). 
By construction, it contains only irreducible monomials 
with respect to ${\cal G}_{i_1\cdots i_n}$, and  it is 
identified with the {\it residue} of  the cut $(i_1\ldots i_n)$.

The integrands $\mathcal{I}_{i_1\cdots i_{\kappa -1}  i_{\kappa+1}  \cdots i_n}$ can be decomposed
repeating the procedure described in  Eqs.~(\ref{Eq:Igen})-(\ref{Eq:DecGenI}). In this case the 
polynomial division of $\mathcal{N}_{i_1\cdots i_{\kappa -1}
  i_{\kappa+1}  \cdots i_n}$ has to be performed 
  modulo the Gr\"obner basis of the ideal 
${\cal J} _{i_1\cdots i_{\kappa -1}   i_{\kappa+1}  \cdots i_n}$, generated by 
the corresponding $(n-1)$ denominators.  The complete multi-pole decomposition 
of the integrand $\mathcal{I}_{i_1\cdots i_n}$  is obtained by successive iterations of 
Eqs.~(\ref{Eq:Igen})-(\ref{Eq:DecGenI}). 

\subsection{Reducibility criterion}
\label{Ssec:Reucibility}
An integrand  $\mathcal{I}_{i_1\cdots i_n} $ is  said to be reducible if it  can be written in terms of lower-point integrands, i.e. 
when the numerator can be written as a linear combination of denominators.
Eqs.~(\ref{Eq:DecGenI}) and (\ref{Eq:fromGtoD}) allow  one to characterize  the reducibility
of the integrands:
\begin{prop}
 The   integrand  $\mathcal{I}_{i_1\cdots i_n} $
 is reducible iff 
 the remainder of the division modulo a Gr\"obner basis vanishes, i.e. iff 
${\cal   N}_{i_1\cdots i_n} \in \mathcal{J}_{i_1\cdots i_n}$. 
\label{Prop:Red}
\end{prop}
A direct consequence of the Proposition~\ref{Prop:Red} is 
\begin{prop}
 \label{Prop:Red4D}
 An integrand   $\mathcal{I}_{i_1\cdots i_n} $ is reducible if the cut $(i_1\cdots i_n)$ leads to a system of equations with no solution.
 \end{prop}
 Indeed  if the system of equations  $D_{i_1} (\zzeta) = \cdots = D_{i_n}(\zzeta)  =0$ 
has no solution, the {\it weak Nullstellensatz} theorem ensures  that 
$1 \in \mathcal{J}_{i_1\cdots i_n}$, i.e. $\mathcal{J}_{i_1\cdots i_n} = P[\zzeta]$. 
Therefore any polynomial in $\zzeta$  is in the ideal. Any numerator function 
$\mathcal{N}_{i_1\cdots i_n}$ is  polynomial in the integration momenta,  thus 
$\mathcal{N}_{i_1\cdots i_n} \in \mathcal{J}_{i_1\cdots i_n} $ and it can 
 be expressed as a combination of the denominators  $D_{i_1} (\zzeta), \ldots,  D_{i_n}(\zzeta)$~\cite{Mastrolia:2012an, Kleiss:2012yv}.
In this case Eq.~(\ref{Eq:DecGen}) becomes
\beq
\mathcal{I}_{i_1\cdots i_n}  =  
 \sum_{\kappa=1}^{n}   \mathcal{I}_{i_1\cdots i_{\kappa -1} i_{\kappa+1} i_n} \ .
\label{eq:RedCrit}
\eeq

\medskip
The reducibility criterion and the recurrence relation~(\ref{Eq:DecGen}) 
are the two mathematical properties underlying the integrand decomposition
of scattering amplitudes, at any order in perturbation theory. 
If the $n$ denominators cannot vanish simultaneously, the corresponding integral is reducible, namely it can be written in terms of integrands with $(n-1)$ denominators. 
If the $n$-ple cut leads to a consistent  system  of equations,  we extract the polynomial form  of the residue as the 
remainder of the division of the numerator  modulo the Gr\"obner  basis associated to the  $n$-ple cut. 
The quotients of the polynomial division generate integrands with $(n-1)$ denominators which should 
undergo the same decomposition.  
The algorithm will stop when all cuts are exhausted, and no denominator is left.
Upon integration, the nonvanishing terms present in each residue may give rise to master integrals.

\medskip
Each residue $\Delta_{i_1\cdots i_n}$ belongs to a vector subspace $Q_{i_1\cdots i_n}[\zzeta]$ of $P[\zzeta]$.
Its dimension is independent  of the choice of the basis.  The residue can be expressed in terms of the 
ISPs writing the components $\zzeta$ in terms of scalar products.  A suitable choice of the bases
of the loop  momenta allow  one to write the ISPs as multivariate monomials  in $\zzeta$ generating 
$Q_{i_1\cdots i_n}[\zzeta]$.  When the number of external legs of the cut  diagram 
is less than five, then the ISPs may involve {\it spurious} terms. 
As in the one-loop 
case~\cite{Ossola:2006us,Ossola:2007bb,Ossola:2007ax,Ossola:2008xq,Ellis:2007br,Giele:2008ve,Ellis:2008ir,Mastrolia:2008jb, Mastrolia:2012bu,Mastrolia:2011pr,Badger:2012dp}, they originate from 
the components of the loop momenta
belonging to the orthogonal space, i.e.\ the space orthogonal to the one spanned by
the independent external momenta of the cut  diagram.

\subsection{Maximum-cut Theorem}
Following~\cite{Mastrolia:2012an}, we define  \emph{Maximum cut} an $n$-ple cut $D_{i_1} (\zzeta) = \cdots = D_{i_n}(\zzeta)  =0$  fully constraining 
all the components $\zzeta$ of the loop momenta.  Examples of maximum cuts for the one-loop case are the $4$-ple  ($5$-ple) cut in $4$ ($d=4-2\epsilon$) dimensions. We assume that, at 
nonexceptional phase-space points,  a  maximum cut  has a finite number $n_s$ 
of solutions, each with multiplicity one.   
Under these assumptions one can prove the following theorem~\cite{Mastrolia:2012an}

\begin{theo}[Maximum cut]
The residue at the
maximum cut is a polynomial paramatrised by $n_s$ coefficients,
which admits a univariate representation of degree
$(n_s - 1)$.
\label{PropM}
\end{theo}
The  maximum-cut theorem  guarantees  that the maximum number of terms needed to parametrize the residue of the maximum cut is exactly equal to $n_s$.
Therefore it guarantees the full reconstruction of  the residue  by sampling the integrand on the  
$n_s$ solutions of the maximum cut. Theorem~\ref{PropM}   generalizes at any loop the simplicity of the one-loop 
maximum  cuts~\cite{Britto:2004nc,Ossola:2006us}.
Indeed, in $d = 4 - 2 \epsilon$ dimensions, the residue of the quintuple cut
is parametrized by one coefficient,
and can be reconstructed by sampling on the single solution of the cut itself.
Similarly the two
coefficients of the residue of the quadruple cut in four dimensions can be determined
by sampling the integrand on the two solutions of the cut.

\subsection{Two-loop integrand reduction}
\label{sec:Reduction}
In four dimensions, the  generic two-loop  $n$-denominator integral ${\cal A}_n$  reads as follows
\bea
 {\cal A}_n &=& 
\int d^{4} q \int d^{4}  k \ 
 \mathcal{I}_n ( q,   k)  
= \int d^{4} q \int d^{4}  k \ 
\frac{{\cal N}_{1\cdots n}( q,  k)
}{
\db{1}\ \db{2}\ \cdots \db{n}
} \ , \nn
 \db{i} &=& (\alpha_{1,i} q + \alpha_{2,i}   k + p_i)^2-m_i^2
\label{def:An2loop}
\eea
Every integrand with more than eight denominators $\db{i}$ leads to a system of equations with no solution for its cut\footnote{A potential ambiguity may arise in topologies with nine denominators two of which are degenerate.  However in this case the one-loop subtopology contains at least six denominators yielding thus a system of equations with no solution.}.
Proposition~\ref{Prop:Red4D} implies that such an integrand is reducible and can therefore be expressed in terms of integrands with eight or less denominators. The recursive procedure 
described in Section~\ref{Ssec:Reucibility} leads to the following multipole decomposition 
\bea
\mathcal{I}_n 
&=& 
\sum_{i_1  < \!<  i_8=1}^n {\Delta_{i_1 \cdots i_8} \over \db{i_1} \cdots \db{i_8}} + 
\sum_{i_1  < \!<  i_7=1}^n {\Delta_{i_1 \cdots i_7} \over \db{i_1} \cdots \db{i_7}} + 
\cdots +
\sum_{i_1 < i_2=1}^n {\Delta_{i_1 i_2} \over \db{i_1} \db{i_2}} 
+ \sum_{i = 1}^n {\Delta_{i} \over \db{i}} + {\cal Q}_{\varnothing} \; , \qquad
\label{eq:2loop:gendeco}
\eea
where $ i_a < \!< i_b $ stands for a lexicographic order
$i_a < i_{a+1} < \ldots < i_{b-1} < i_b$.
Equivalently,  the numerator decomposition formula reads
\bea
\mathcal{N}_{1 \cdots n }
&=& 
\sum_{i_1< \!<  i_8=1}^n {\Delta_{i_1 \cdots i_8}} \prod_{j \ne i_1, \ldots, i_8}^n \db{j} +
\sum_{i_1 < \!<   i_7=1}^n {\Delta_{i_1 \cdots i_7}} \prod_{j \ne i_1, \ldots, i_7}^n \db{j}+
\cdots  \nn & & 
+ \sum_{i_1 < i_2=1}^n {\Delta_{i_1 i_2}} \prod_{j \ne i_1,i_2}^n \db{j}
+ \sum_{i = 1}^n {\Delta_{i}} \prod_{j \ne i}^n \db{j} + 
{\cal Q}_{\varnothing} \prod_{j=1}^n \db{j}
\ . \qquad
\eea
The residue $\Delta_{i_1\cdots i_k}$ is obtained from the corresponding rank $r_{i_1\cdots i_k}$
integrand $\mathcal{I}_{i_1\cdots i_k}$ using the following procedure:
\begin{enumerate}
\item Decompose the loop momenta in using two bases  $\{\tau_i\}_{i=1,\ldots,4}$ and $\{e_j\}_{j=1,\ldots,4}$:
\bea
q^\mu = -p^\mu_0 + \sum_{i=1}^4 x_i \; \tau^\mu_i , \qquad k^\mu = -r^\mu_0 + \sum_{i=1}^4 y_j \; e^\mu_j \; .
\label{Eq:subQK}
\eea
In this case $\zzeta \equiv (y_1,\ldots,y_4,x_1,\ldots,x_4)$.
\item Consider a generic rank $r_{i_1\cdots i_k}$ polynomial in $\zzeta$
\begin{align}
\mathcal{N}_{i_1 \cdots i_k}(\zzeta) &= \sum_{\vec j \in J(r_{i_1\cdots i_k})}  \alpha_{\vec j } \left ( \prod_{i=1}^{8} z_i^{j_i}  \right ), &
 J(r)  \equiv \{\vec j \in \mathbb{N}^8 :  \sum_{i=1}^8  j_i \le r \}  \,  .  
\end{align}
\item Choose a monomial order and  construct a Gr\"obner basis $\mathcal{G}_{i_1 \cdots  i_k}=\{g_{1}(\zzeta), \ldots  , g_{\emm}(\zzeta) \}$, 
generating the ideal ${\cal J}_{i_1 \cdots i_k} = \langle D_{i_1}, \ldots, D_{i_k}\rangle$.
\item Divide  $\mathcal{N}_{i_1\cdots i_k}$ modulo 
$\mathcal{G}_{i_1 \cdots i_k}$ holding the remainder $\Delta_{i_1\cdots i_k}$.
\end{enumerate}
The integrand decomposition (\ref{eq:2loop:gendeco})
allows one to express the amplitude in
terms of MIs, associated to diagrams with $8$, $7$,$\ldots$ , $2$ denominators. 
Depending on the powers of the integration momenta appearing in the numerator, 
the multivariate division may also generate the single-cut residues $\Delta_i$,
and the quotients of the last divisions,  ${\cal Q}_{\varnothing}$. 
These two contributions generate spurious terms only but they are needed for the complete reconstruction of the integrand. 
The term  ${\cal Q}_{\varnothing}$ is  non-cut-constructible: its 
determination   requires to sample 
the numerator away from the solutions of the
multiple cuts.

We expect that  the integrand-reduction formula could be extended to $d$ dimensions, where 
additional degrees of freedom related to $\vec \lambda_q$ and $\vec \lambda_k$ enter~\cite{Boughezal:2011br}.

\def\lp{\left(}
\def\rp{\right)}

\def\A{\mathcal{A}}
\def\S{\mathcal{S}}
\def\O{\mathcal{O}}
\def\N{\mathcal{N}}
\def\nn{\nonumber \\}

\def\la{\langle}
\def\ra{\rangle}
\def\lb{[}
\def\rb{]}
\def\spa#1{\langle#1\rangle}
\def\spb#1{[#1]}
\def\spab#1{\langle#1]}
\def\spba#1{[#1\rangle}

\def\hi{\hat{i}}
\def\hj{\hat{j}}
\def\hQ{\hat{Q}}

\def\vi{u_1}
\def\wu{u_2}
\def\es{u_3}
\def\mo{w}

\def\ETA{\eta}
\def\SIG{\sigma}

\begin{figure}[t]
\begin{center}
\includegraphics[scale=1.] {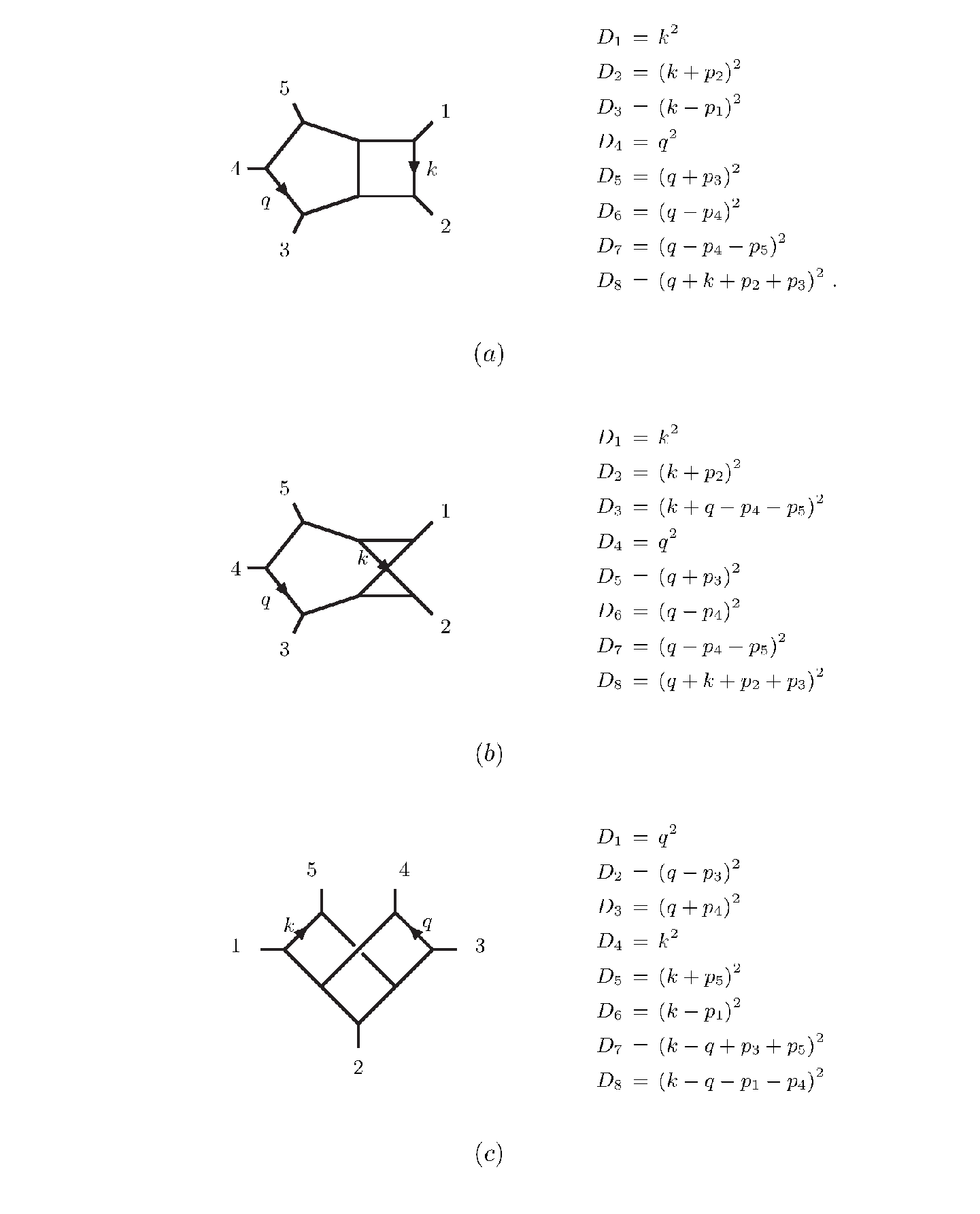}
\end{center}
\caption{Five-point diagrams entering the amplitudes in $\mathcal{N}=4$ SYM and $\mathcal{N}=8$ SUGRA.  They 
are the pentabox diagram $(a)$, the crossed  pentabox diagram $(b)$ and the double pentagon diagram $(c)$.
For each diagram, the definition of the denominators is shown as well.}
\label{fig:diagram}
\end{figure}

\section{Five-point amplitudes in ${\cal N}=4$ SYM}
The  five-point amplitude in ${\cal N}=4$ SYM can be expressed in terms of 
six diagrams~\cite{Carrasco:2011mn}.  The color ordered   amplitude 
 is given by a  sum over  the cyclic permutations of the external momenta. 
We apply the integrand reduction only to the three diagrams  depicted in 
Fig.~\ref{fig:diagram}. The other three diagrams are trivially expressed in terms 
of scalar integrals,  since  their numerator  is independent of the loop momenta.
We consider one integrand at a time  and we obtain its decomposition by evaluating its numerator 
on solutions of multiple cuts, i.e.  on  values of the loop momenta such that some of 
its denominators vanish.  \\

We introduce the notation $\mathcal{I}^{(4, i)}$ \
( $\mathcal{N}^{(4, i)}$)
to denote the integrand (numerator)  of the diagram in Fig.~\ref{fig:diagram} $(i)$  in 
$\mathcal{N} = 4$ SYM. 
The integrand of diagrams in Fig.~\ref{fig:diagram}  are
\begin{align}
\label{Eq:pentaboxN4}
& {\cal I}^{(4,a)}_{1\cdots 8} ( q, k ) =  \frac{{\cal N}^{(4,a)}_{1\cdots 8} ( q, k ) }{\db{1} \cdots \db{8}}\; ,  &  {\cal N}^{(4,a)}_{1\cdots 8} ( q, k ) &= 2\, q \cdot \vi + \beta_1 \; , \pagebreak[0]    \\
\label{Eq:pentacrossN4}
& {\cal I}^{(4,b)}_{1\cdots 8} ( q, k ) =  \frac{{\cal N}^{(4,b)}_{1\cdots 8} ( q, k ) }{\db{1} \cdots \db{8}}\; ,   &  {\cal N}^{(4,b)}_{1\cdots 8} ( q, k ) &= 2\, q \cdot \vi + \beta_1 \; , \pagebreak[0]   \\
 \label{eq:5pwn4symnumpar}
& {\cal I}^{(4,c)}_{1\cdots 8} ( q, k ) =  \frac{{\cal N}^{(4,c)}_{1\cdots 8} ( q, k ) }{\db{1} \cdots \db{8}}\; ,   &    \mathcal{N}^{(4,c)}_{1\cdots 8}(q,k)  &= 2\, q \cdot \wu  + 2\, k \cdot \es + \beta_2 + \beta_3 \pagebreak[0]  
\; ,
\end{align}
The vectors $\vi^\mu$, $\wu^\mu$, and $\es^\mu$ and the constants $\beta_i$
are defined as~\cite{Carrasco:2011mn}
\begin{align}
\label{eq:valphan4pbv}
  \vi^\mu =& \frac{1}{4}\Big( \gamma_{35  124}( p_5^\mu-p_3^\mu )+
\gamma_{34 125}( p_4^\mu-p_3^\mu)+\gamma_{45 123}( p_5^\mu-p_4 ^\mu )+2\,
\gamma_{12 345}( p_2^\mu-p_1^\mu )  \Big)  \; , \quad \displaybreak[0]    \\
\label{eq:valphan4ptv}
  \wu^\mu  =& \frac{1}{4}\Big(
\gamma_{23145}(p_2^\mu-p_3^\mu)+\gamma_{24135}(p_2^\mu-p_4^\mu)+\gamma_{34125}(p_3^\mu-p_4^\mu)+2\,
\gamma_{15234}(p_1^\mu-p_5^\mu) \Big) \; ,  \displaybreak[0]  \\
%
\label{eq:valphan4plv}
  \es^\mu  =& \frac{1}{4}\Big(
\gamma_{12345}(p_1^\mu-p_2^\mu)+\gamma_{25134}(p_2^\mu-p_5^\mu)+\gamma_{15234}(p_1^\mu-p_5^\mu)+2\,
\gamma_{34125}(p_3^\mu-p_4^\mu) \Big) \; , \displaybreak[0]   \\
%
  \beta_1   =& \frac{1}{4}\Big( \gamma_{35 124}(s_{34}+s_{12}+s_{35} ) + 2\,
\gamma_{34 125}\, s_{12}+\gamma_{45 123}( s_{34}+s_{12}+s_{35} )  \displaybreak[0] \nn
\label{eq:valphan4pbalpha}
   +&2\, \gamma_{12 345}( s_{23}-s_{13} )  \Big) \;  ,  \displaybreak[0]  \\
%
\label{Eq:beta2}
  \beta_2  = & \frac{1}{4}\Big(  -(\gamma_{23145}+\gamma_{24135})\,
s_{23}+\gamma_{34125}(s_{15}+s_{34}+2\, s_{23})-2\,
\gamma_{15234}(s_{13}-s_{35}) \Big) \; ,   \displaybreak[0]  \\
%
  \beta_3  = & \frac{1}{4}\Big(  (\gamma_{12345}-\gamma_{25134})\,
s_{12}+\gamma_{15234}(s_{34}+s_{15}+2\, s_{12})-2\,
\gamma_{34125}(s_{13}-s_{14}) \Big) \; .
\end{align}
where the kinematic invariants $s_{ij}$ and the functions $\gamma$ read as follows
\begin{align}
  s_{ij} & \equiv (p_i+p_j)^2 = 2\, (p_i\cdot p_j) \\
  \label{eq:carrascogamma12345}
  \gamma_{12 345} & \equiv \left ( \frac{[1\, 2] [2\, 3] [3\, 4]
    [4\, 5] [5\, 1]}{[1\, 4]
    [2\, 3] \langle 1\, 2\rangle  \langle
    3\, 4\rangle -[1\, 2] [3\, 4]
    \langle 1\, 4\rangle  \langle 2\, 3\rangle }  \right )- \left ( 1 \leftrightarrow 2\right )  . 
\end{align}

In ${\cal N}=4$ SYM, given the simple form of the numerators, the multipole decomposition of the integrands only requires one
iteration.  The numerators can be decomposed as
\bea
\label{Eq:N4a}
\label{Eq:N4b}
\label{Eq:N4c}
{\cal N}^{(4,x)}_{1\cdots 8} ( q, k ) &=&  \Delta_{12345678}  +  \sum_{i=1}^7  \Delta_{1\cdots (i-1)(i+1) \cdots 8} \db{i} \,\,\, , \,\,\, x=a,b,c \; . 
\eea
The number of  $7$-ple residues of the integrands ${\cal N}^{(4,a)}$ and ${\cal N}^{(4,b)}$ is almost halved since the numerator
depends on $q$ only, thus $\Delta_{1\cdots (i-1)(i+1) \cdots 8} =0$ for $i \neq 4,5,6,7$. 

In the next subsections,  we list  the parametrization of the residues entering our computation, namely of the residues in Eq.~(\ref{Eq:N4a}). All the eightfold residues are related to {\it maximum cuts}. According to the 
maximum-cut theorem~\cite{Mastrolia:2012an},  the number of coefficients needed to parametrize the residue of a maximum cut is finite 
and equal to the number of the solutions of the corresponding cut.  We found the most general parametrization of the eightfold 
residue, which is process-independent and valid for numerators of any rank in both $q$ and $k$.
The parametrization of the sevenfold residues  is also process-independent and it is given for the case of renormalizable numerators of rank six at most, which is more than we need for the applications presented in this paper.  The parametrization of 
higher rank numerators can be obtained  including   additional terms with higher powers  of the loop momenta  in the residues~\cite{Mastrolia:2012bu}. 
The number of coefficients required to parametrize the residues agrees with an independent computation performed using a technique based 
on Gram determinants.

\begin{figure}[t]
\begin{center}
\includegraphics[scale=1.] {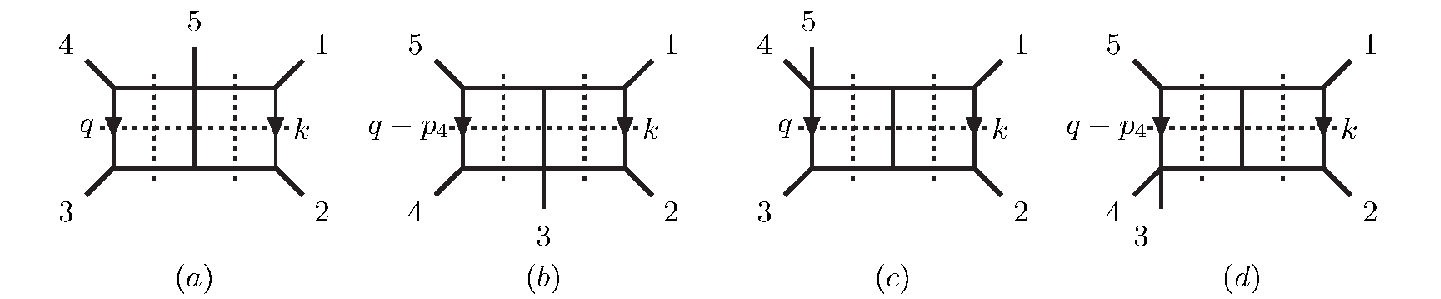}
\end{center}
\caption{Cut diagrams of the sevenfold cuts.  Starting from the left, we show the  diagram of the cut $(1234568)$, $(1234678)$,  
$(1234578)$, and  $(1235678)$.
}
\label{fig:5p:7foldcut:1234568}
\end{figure}

\newcommand{\trule}{\rule[-1.5mm]{0mm}{6mm}}
\TABULAR[t]{l | l | l | l}{
\hline \hline \trule  cut              &  bases                                          & $\zzeta$                                                        & Monomials in the residue  \\   \hline \trule
$(12345678)$     & \small Eq. (\ref{def:ebasis8fold})  & \small $(y_4,y_3,y_2,y_1,x_4,x_3,x_2,x_1)$               &  \small $\mathcal{S}_{12345678} = \{ 1, x_1, y_1, y_2 \} $  \\  \hline \trule
$(1234568)$       &  \small Eq. (\ref{def:ebasis8fold})  &  \small $(y_4,y_3,y_2,y_1,x_4,x_3,x_2,x_1)$                  &  \small $\mathcal{S}_{1234568} =  \{  1,x_1,x_1^2,x_1^3,x_1^4,x_2,x_1 x_2,$ \\
&&&                    \small         $x_1^2 x_2,x_1^3 x_2, y_1,x_1 y_1,x_1^2 y_1,x_1^3 y_1, x_1^4 y_1,$ \\
&&&                    \small         $x_2  y_1,x_1 x_2 y_1, x_1^2 x_2 y_1,x_1^3 x_2 y_1,y_1^2,x_1y_1^2,$ \\
&&&                    \small         $x_2 y_1^2, y_1^3, x_1 y_1^3, x_2 y_1^3,y_1^4,x_1y_1^4,x_2 y_1^4,y_2,$ \\
&&&                   \small           $x_1 y_2,y_1 y_2,y_1^2 y_2,y_1^3 y_2  \}$  \\ \hline \trule
$(1234678)$       &\small Eq.  (\ref{def:ebasis8fold})  & \small$(y_4,y_3,y_2,y_1,x_4,x_3,x_2,x_1)$               &   \small $\mathcal{S}_{1234678} =\mathcal{S}_{1234568}$ \\ \hline \trule
$(1234578)$       & \small Eq. (\ref{def:ebasis7fold1})  &  \small $(y_4,y_3,y_2,y_1,x_4,x_3,x_2,x_1)$              &   \small $\mathcal{S}_{1234578} = \{1,x_2,x_2^2,x_2^3,x_2^4,x_4,x_2 x_4,$ \\
&&& \small      $x_2^2 x_4,x_2^3x_4, y_1,x_2 y_1,x_2^2 y_1,x_2^3 y_1,x_2^4 y_1,x_4y_1,$ \\
&&& \small      $x_2 x_4 y_1,x_2^2 x_4 y_1,x_2^3 x_4 y_1,y_1^2,x_2y_1^2,x_4 y_1^2,y_1^3,x_2 y_1^3,$ \\
&&& \small      $x_4 y_1^3,y_1^4,x_2y_1^4,x_4 y_1^4,y_4,x_2 y_4,y_1 y_4,$\\
&&& \small      $y_1^2 y_4,y_1^3y_4 \}$ \\ \hline \trule
$(1235678)$       &\small Eq.  (\ref{def:wbasis7fold:1235678})  &  \small  $(y_4,y_3,y_2,y_1,x_4,x_3,x_2,x_1)$                &   \small $\mathcal{S}_{1235678} =\{ 1,x_1,x_1^2,x_1^3,x_1^4,x_2,$\\
&&&  \small $x_1 x_2,x_1^2 x_2,x_1^3x_2,y_1,x_1 y_1,x_1^2 y_1,$ \\
&&&  \small $x_1^3 y_1,x_1^4 y_1,x_2 y_1,  x_1 x_2 y_1,x_1^2 x_2 y_1,x_1^3 x_2y_1,$ \\
&&&  \small $y_1^2,x_1 y_1^2,x_2 y_1^2,y_1^3,x_1 y_1^3,x_2 y_1^3,$ \\
&&&  \small $y_1^4,  x_1 y_1^4,x_2 y_1^4,y_3,x_1 y_3,y_1 y_3,y_1^2 y_3,y_1^3 y_3 \}$ \\    \hline \hline 
}
{Set of monomials parametrizing the residues entering the decomposition of the five-point pentabox diagram.  They have all been found using degree lexicographic monomial ordering.  For each cut 
the bases and the chosen ordering for loop variables are shown as well.
\label{Tab:Pentabox}}

\subsection{Residue of the planar pentabox }
\label{sec:Pentabox}
The decomposition of the pentabox diagram in Fig.~\ref{fig:diagram} $(a)$ 
requires the parametrization of the residues of the eightfold cut $(12345678)$ and the 
sevenfold cuts depicted in Fig.~\ref{fig:5p:7foldcut:1234568}.    The parametrization is obtained 
using the procedure  described in Section~\ref{sec:ReductionG}. 
The relevant  bases are
\bea
\label{def:ebasis8fold} 
&&\begin{cases}
\begin{aligned}
 r_0^\mu  &= 0^\mu,   &  e^\mu_1      &= p_3^\mu, &   e^\mu_2      &= p_4^\mu, & e_3^\mu     &= \frac{\langle 3|\gamma^\mu |4 ]}{2} , & e_4^\mu &= \frac{\langle 4|\gamma^\mu | 3 ]}{2} ,  \quad  \\
 p_0^\mu  &=0^\mu,    &   \tau^\mu_1 &= p_2^\mu, &   \tau^\mu_2 &= p_1^\mu, & \tau_3^\mu &= \frac{\langle 2|\gamma^\mu | 1 ]}{2} , & \tau_4^\mu &= \frac{\langle 1 |\gamma^\mu |2 ]}{2} , \quad
 \end{aligned} \\[5.0ex]
  x_1 = \frac{(q\cdot p_1)}{(p_1 \cdot p_2)}\ , \qquad 
x_2 = \frac{(q\cdot p_2)}{(p_1 \cdot p_2)}\ , \qquad 
y_1 = \frac{(k\cdot p_4)}{(p_3 \cdot p_4)}\ , \qquad 
y_2 = \frac{(k\cdot p_3)}{(p_3 \cdot p_4)}\ ;  
\end{cases}\pagebreak[1]  \\[2.0ex]
\label{def:ebasis7fold1}
&&\begin{cases}
\begin{aligned}
 r_0^\mu  &= 0^\mu,   &  e^\mu_1      &= p_1^\mu, &   e^\mu_2      &= p_3^\mu, & e_{3,4}^\mu     &=  \frac{
 \langle 3| 2 |1]
 \langle 1|\gamma^\mu | 3] \pm   \langle 1 |2 |3]
 \langle 3|\gamma^\mu | 1] 
 }{4}    , \quad  \\
p_0^\mu  &=0^\mu,    &   \tau^\mu_1 &= p_1^\mu, &     \tau^\mu_2 &= p_3^\mu, &     \tau^\mu_{3,4} &= \frac{
 \langle  3| 2 | 1]
 \langle  1|\gamma^\mu | 3] \pm   \langle  1| 2 | 3]
 \langle  3|\gamma^\mu | 1] 
 }{4}  ,  \quad 
  \end{aligned} \\[5.0ex]
 x_1 =\frac{ (q\cdot p_3)}{(p_1\cdot p_3)}, \qquad 
x_4 = \frac{(q\cdot \tau_4)}{\tau_4^2}, \qquad 
y_1 = \frac{(k\cdot p_3)}{(p_1\cdot p_3)}, \qquad 
y_4 = \frac{(k\cdot e_4)}{e_4^2} ;
\end{cases} \pagebreak[1]  \\[2.0ex]
&&\begin{cases}
\begin{aligned}
 r_0^\mu  &= 0^\mu,   &  e^\mu_{1,4}      &=  \frac{\langle 3|2 |1]
 \langle 1|\gamma^\mu |3] \mp   \langle 1|2 |3]
 \langle 3|\gamma^\mu |1] 
 }{4}  & e^\mu_2 &=  p_3^\mu, &   e^\mu_3      &= p_1^\mu  ,  \\ 
p_0^\mu  &=-p_4^\mu,    &   \tau^\mu_{1,4} &=\frac{\langle 3|2 |1]
 \langle 1|\gamma^\mu |3] \mp   \langle 1|2 |3]
 \langle 3|\gamma^\mu |1] 
 }{4}, &     \tau^\mu_2 &= p_3^\mu, &     \tau^\mu_{3} &= p_{1}^\mu ,
\end{aligned} \\[5.0ex]
 x_1 =\frac{ ((q-p_4) \cdot e_1)}{e_1^2}, \qquad 
x_2 = \frac{((q-p_4) \cdot p_1)}{(p_1\cdot p_3)}, \qquad 
y_1 = \frac{(k\cdot \tau_1)}{\tau_1^2}, \qquad 
y_3 = \frac{(k\cdot p_3)}{(p_1\cdot p_3)} . \qquad   
\end{cases}  \pagebreak[1] 
\label{def:wbasis7fold:1235678}
\eea
The residue of  each cut $(i_1\ldots i_\kappa)$ is written in terms of  a set of monomials
 $\mathcal{S}_{i_1 \cdots i_\kappa}$, collected in Table~\ref{Tab:Pentabox}.

\begin{figure}[]
\vspace*{1.5cm}
\begin{center}
\includegraphics[scale=1.] {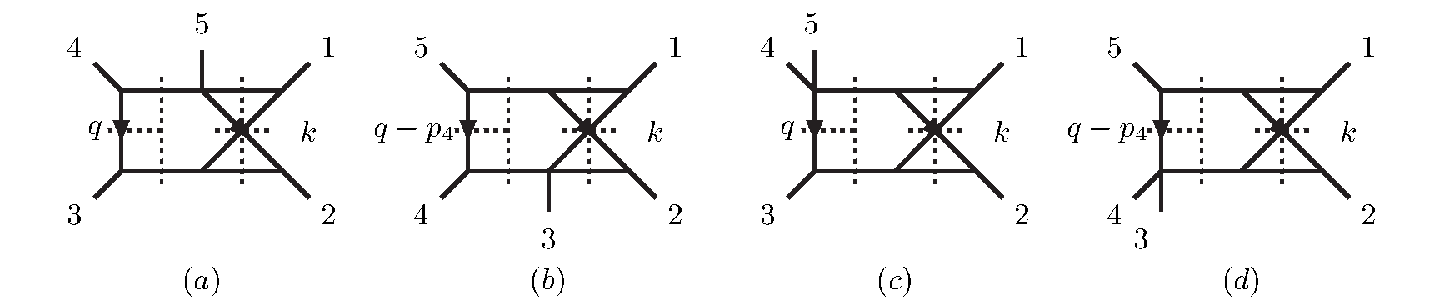}
\end{center}
\caption{Cut diagrams of the sevenfold cuts.  Starting from the left, we show the  diagram of the cut $(1234568)$, $(1234678)$,  
$(1234578)$, and  $(1235678)$.}
\label{fig:5p:7foldcut:cross:1234568}
\end{figure}

\TABULAR[t]{l | l | l | l}{
\hline \hline \trule  cut              &  bases                                          & $\zzeta$                                                        & Monomials in the residue  \\   \hline \trule
$(12345678)$    & \small Eq. (\ref{def:ebasis8fold})  & \small $(x_4,x_3,x_2,y_3,y_4,x_1,y_2,y_1)$                &  \small $\mathcal{S}_{12345678} = \{ 1, x_1, y_1, y_2 \} $  \\  \hline \trule
$(1234568)$      &  \small Eq. (\ref{def:ebasis8fold})  &  \small  $(y_4,y_3,y_2,y_1,x_4,x_3,x_2,x_1)$                 &  \small $\mathcal{S}_{1234568} =  \{1,x_1,x_1^2,x_1^3,x_1^4,x_1^5,x_1^6,x_2,$\\
&&&  \small $x_1 x_2,x_1^2  x_2,x_1^3 x_2,x_1^4 x_2,x_1^5 x_2,$\\
&&& \small $y_1,x_1 y_1,x_1^2   y_1,x_1^3 y_1,x_1^4 y_1,x_1^5 y_1,x_2 y_1,$\\
&&& \small $x_1 x_2   y_1,x_1^2 x_2 y_1,x_1^3 x_2 y_1,x_1^4 x_2 y_1,y_1^2,x_1y_1^2,$\\
&&& \small $x_2 y_1^2,y_1^3,x_1 y_1^3,x_2 y_1^3,y_1^4,x_1 y_1^4,x_2 y_1^4,y_2,$\\
&&& \small $x_1 y_2, y_1 y_2,y_1^2 y_2,y_1^3   y_2 \}$ \\  \hline \trule
$(1234678)$     &\small Eq.  (\ref{def:ebasis8fold})  & \small  $(y_4,y_3,y_2,y_1,x_4,x_3,x_2,x_1)$                 &   \small $\mathcal{S}_{1234678} =\mathcal{S}_{1234568}$ \\ \hline \trule
$(1234578)$       & \small Eq. (\ref{def:ebasis7fold1})  &  \small  $(y_4,y_3,y_2,y_1,x_4,x_3,x_2,x_1)$                 &   \small $\mathcal{S}_{1234578} = \{1,x_2,x_2^2,x_2^3,x_2^4,x_2^5,x_2^6,x_4,$\\
&&&   \small  $x_2 x_4,x_2^2 x_4,x_2^3 x_4,x_2^4 x_4,x_2^5 x_4,y_1,x_2 y_1,$\\
&&&  \small   $x_2^2 y_1,x_2^3 y_1,x_2^4   y_1,x_2^5 y_1,x_4 y_1,x_2 x_4 y_1,x_2^2 x_4 y_1,$ \\
&&&   \small $x_2^3 x_4 y_1,x_2^4 x_4 y_1,y_1^2,x_2 y_1^2,x_4 y_1^2,y_1^3,x_2 y_1^3,$\\
&&&  \small  $x_4 y_1^3,y_1^4,x_2 y_1^4,x_4 y_1^4,y_4,x_2 y_4,y_1y_4,y_1^2 y_4,y_1^3 y_4 \}$ \\ \hline \trule
$(1235678)$       &\small Eq.  (\ref{def:wbasis7fold:1235678})  &  \small  $(y_4,y_2,y_3,y_1,x_4,x_3,x_2,x_1)$                 &   \small $\mathcal{S}_{1235678} = \{1,x_1,x_1^2,x_1^3,x_1^4,x_1^5,x_1^6,x_2,$\\
&&& \small $x_1 x_2,x_1^2  x_2,x_1^3 x_2,x_1^4 x_2,x_1^5 x_2,y_1,$\\
&&& \small $x_1 y_1,x_1^2   y_1,x_2 y_1,x_1 x_2 y_1,y_1^2,x_1 y_1^2,x_1^2   y_1^2,$\\
&&& \small $x_2 y_1^2,x_1 x_2 y_1^2,y_1^3,x_1 y_1^3,x_1^2y_1^3,x_2 y_1^3,x_1 x_2 y_1^3,$\\
&&& \small $y_1^4,x_1 y_1^4,x_1^2   y_1^4,x_2 y_1^4,x_1 x_2 y_1^4,y_3,x_1 y_3,$\\
&&& \small $y_1   y_3,y_1^2 y_3,y_1^3 y_3\}$ \\ \hline \hline
}
{The same as Table~\ref{Tab:Pentabox}, but for the five-point crossed pentabox diagram. \label{Tab:Pentacross}}

\subsection{Residue of the  crossed pentabox }
\label{sec:Pentacross}
The diagram in Fig.~\ref{fig:diagram} $(b)$ is decomposed in terms of the residue of the eightfold cut $(12345678)$ and of 
the residue of the  sevenfold cuts  in Fig.~\ref{fig:5p:7foldcut:cross:1234568}.   Each residue can be expressed in terms 
of a set of monomials, as shown in Table~\ref{Tab:Pentacross}. The parametrization is   obtained using the 
multivariate polynomial division described in  Section~\ref{sec:ReductionG}.

\TABULAR[t]{l | l | l | l}{
\hline \hline \trule  cut              &  bases                                          & $\zzeta$                                                        & Monomials in the residue  \\   \hline \trule
$(12345678)$    & \small Eq. (\ref{def:ebasis7foldtwist0})  & \small $(y_4,y_3,y_2,y_1,x_4,x_3,x_1,x_2)$                &  \small $\mathcal{S}_{12345678} = \{1,y_1,x_2,y_1 x_2,x_2^2,x_2^3,x_1,x_2 x_1 \}$  \\  \hline \trule
$(1345678)$      &  \small Eq. (\ref{def:ebasis7foldtwist0})  &  \small  $(x_4,x_3,x_2,x_1,y_4,y_3,y_2,y_1)$                 &  \small $\mathcal{S}_{1345678} =  \{1,y_1,y_1^2,y_1^3,y_1^4,y_1^5,y_1^6,y_2,y_1y_2,$ \\
&&& \small $y_1^2y_2,y_1^3 y_2,y_1^4 y_2,y_1^5 y_2,x_1,$\\
&&& \small $y_1 x_1,y_1^2   x_1,y_1^3 x_1,y_1^4 x_1,y_1^5 x_1,$ \\
&&& \small $y_2 x_1,y_1 y_2 x_1,y_1^2  y_2 x_1, y_1^3 y_2 x_1,$\\
&&& \small $y_1^4 y_2 x_1,x_1^2,y_1 x_1^2,y_2 x_1^2,x_1^3,y_1 x_1^3,$\\
&&& \small $y_2 x_1^3,x_1^4,y_1 x_1^4,y_2 x_1^4,$ \\
&&& \small $x_2,y_1 x_2,x_1 x_2,x_1^2 x_2,x_1^3 x_2\}$\\ \hline \trule
$(1245678)$     &\small Eq.  (\ref{def:ebasis7foldtwist0})  & \small  $(x_4,x_3,x_2,x_1,y_4,y_3,y_2,y_1)$                 &   \small $\mathcal{S}_{1245678} =\mathcal{S}_{1345678}$ \\ \hline \trule
$(2345678)$       & \small Eq. (\ref{def:ebasis7foldtwist1})  &  \small  $(x_1,x_3,x_2,x_4,y_3,y_4,y_2,y_1)$   &   \small $\mathcal{S}_{2345678} =\{1,y_1,y_1^2,y_1^3,y_1^4,y_1^5,y_1^6,y_4,y_1y_4,$ \\
&&& \small $y_1^2 y_4,y_1^3y_4,y_1^4 y_4, y_1^5 y_4,x_2,y_1 x_2,$\\
&&& \small $x_4,y_1 x_4,y_1^2 x_4,y_1^3 x_4,y_1^4 x_4,y_1^5 x_4,$\\
&&& \small $y_4 x_4,y_1 y_4 x_4,y_1^2   y_4 x_4,y_1^3 y_4 x_4,$\\
&&& \small $y_1^4 y_4 x_4,x_2 x_4,x_4^2,y_1   x_4^2,y_4 x_4^2,$\\
&&& \small $x_2 x_4^2,x_4^3,y_1 x_4^3,y_4 x_4^3,x_2 x_4^3,x_4^4,y_1 x_4^4,y_4 x_4^4\Big\} \; , $\\ \hline \trule
$(1234567)$       &\small Eq.  (\ref{def:ebasis7foldtwist0})   &  \small  $(x_4,x_3,x_2,x_1,y_4,y_3,y_2,y_1)$   &   \small $\mathcal{S}_{123567} = \{1,y_1,y_1^2,y_1^3,y_1^4,y_2,y_1 y_2,y_1^2y_2,y_1^3y_2,$\\
&&& \small$x_1,y_1 x_1,y_1^2 x_1,y_1^3 x_1,y_1^4 x_1,y_2 x_1,y_1   y_2 x_1,$\\
&&& \small$y_1^2 y_2 x_1,y_1^3 y_2 x_1,x_1^2,y_1 x_1^2,y_2x_1^2,x_1^3,$\\
&&& \small$y_1 x_1^3,y_2 x_1^3,x_1^4,y_1 x_1^4,y_2   x_1^4,x_2,y_1 x_2,$\\
&&& \small$x_1 x_2,x_1^2 x_2,x_1^3 x_2\}$ \\ \hline \hline
}
{The same as Table~\ref{Tab:Pentabox}, but for the five-point double pentagon diagram in Fig.~\ref{fig:diagram} $(c)$. \label{Tab:Pentatwist}}

\subsection{Residue of the double pentagon }
\label{sec:Pentatwist}
\begin{figure}[]
\begin{center}
\includegraphics[scale=1.] {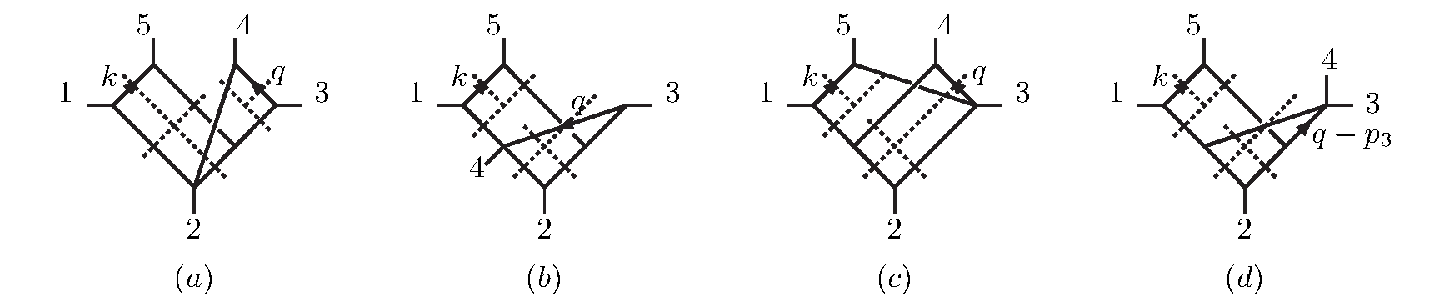}
\end{center}
\caption{Cut diagrams of the sevenfold cuts.  Starting from the left, we show the  diagram of the cut $(1234567)$, $(1245678)$,  
$(1345678)$, and  $(2345678)$.}
\label{fig:7foldcuts:twist}
\end{figure}

The decomposition of the double pentagon diagram requires the parametrization of the residues of the eightfold cut $(12345678)$ and all the sevenfold cuts.
However, the topology in Fig.~\ref{fig:diagram} $(c)$ is invariant under the transformation 
\bea
p^\mu_1 \leftrightarrow p^\mu_3,\qquad  p^\mu_4 \leftrightarrow p^\mu_5,\qquad q^\mu  \leftrightarrow k^\mu \, ;
\label{Eq:transform}
\eea 
thus the only sevenfold cut needed are   $(1345678)$, $(1245678)$, $(2345678)$, and $(1234567)$, depicted in 
Fig.~\ref{fig:7foldcuts:twist}. 
The remaining sevenfold cuts can be obtained using the  transformation~(\ref{Eq:transform}).  The eightfold
cut is a maximum cut. It exhibits eight solutions and it is parametrized by eight coefficients, in accordance with the maxim-cut theorem.

The sets of monomials  parametrizing  the relevant residues are collected in  Table~\ref{Tab:Pentatwist}.  They
are obtained by multivariate polynomial division using the following bases  
%
\bea
\label{def:ebasis7foldtwist0} 
&&\begin{cases}
\begin{aligned}
  r_0^\mu  &=0^\mu,    &   e^\mu_1 &= p_4^\mu, &   e^\mu_2 &= p_3^\mu, & e_3^\mu &= \frac{\langle 4|\gamma^\mu | 3 ]}{2} , & e_4^\mu &= \frac{\langle 3 |\gamma^\mu |4 ]}{2} , \quad \\
 p_0^\mu  &= 0^\mu,   &  \tau^\mu_1      &= p_5^\mu, &   \tau^\mu_2      &= p_1^\mu, & \tau_3^\mu     &= \frac{\langle 5|\gamma^\mu |1 ]}{2} , & \tau_4^\mu &= \frac{\langle 1|\gamma^\mu | 5 ]}{2} ,  \quad
 \end{aligned} \\[5.0ex]
x_1 = \frac{(q\cdot p_1)}{(p_5 \cdot p_1)}\ , \qquad 
x_2 = \frac{(q\cdot p_5)}{(p_5 \cdot p_1)}\ , \qquad   
  y_1 = \frac{(k\cdot p_3)}{(p_3 \cdot p_4)}\ , \qquad 
y_2 = \frac{(k\cdot p_4)}{(p_3 \cdot p_4)}\ ;
\end{cases}\pagebreak[1]  \\[2.0ex]
&&\begin{cases}
\begin{aligned}
r_0^\mu  &=0^\mu,    &   e^\mu_1 &= p_1^\mu, &     e^\mu_2 &= p_3^\mu, &     e^\mu_{3,4} &= \frac{
 \langle 1|4 |3]
 \langle 3|\gamma^\mu |1] \pm   \langle 3|4 |1]
 \langle 1|\gamma^\mu |3] 
 }{4}   , \\
 p_0^\mu  &= -p_3^\mu,   &  \tau^\mu_1      &= p_1^\mu, &   \tau^\mu_2      &= p_3^\mu, & \tau_{3,4}^\mu     &=  \frac{
 \langle 1|4 |3]
 \langle 3|\gamma^\mu |1] \pm   \langle 3|4 |1]
 \langle 1|\gamma^\mu |3] 
 }{4}   ,  \\ 
 \end{aligned} \\[5.0ex]
x_2 = \frac{((q-p_3)\cdot p_1)}{(p_1\cdot p_3)}, \quad 
x_4 = \frac{((q-p_3)\cdot \tau_4)}{\tau_4^2}, \quad  
 y_1 =\frac{ (k\cdot p_3)}{(p_1\cdot p_3)}, \quad 
y_4 = \frac{(k\cdot e_4)}{e_4^2}.
\end{cases}
\label{def:ebasis7foldtwist1}
\eea

\section{Seminumerical integrand reduction}
\label{sec:Numerics4}
In the previous section we illustrated how to determine the general structure of the residues by means of the multivariate polynomial division.
Knowing this structure, we can proceed and numerically perform the integrand reduction to extract the values of all process-dependent coefficients which appear in the residues.
The decomposition can be checked by verifying the identity between the original  numerator and its reconstruction,
i.e. between l.h.s. and r.h.s. of Eq.~(\ref{Eq:N4a}), for  arbitrary values of the integration 
momenta $q$ and $k$. This procedure is known as global $N = N$ test of the integrand reduction.

\subsection{Planar pentabox diagram}
 \label{SSec:NumPentabox}
\paragraph{Eightfold cut.}
The residue of the eightfold cut can be parametrized using the monomials in Table~\ref{Tab:Pentabox}: 
\begin{equation}
\Delta_{12345678} =  \ci{12345678}{0} +  \ci{12345678}{1} (q\cdot p_1) +
 \ci{12345678}{2} (k\cdot p_3) 
  +  \ci{12345678}{3} (k\cdot p_4)  \, .
 \label{Eq:DeltaN4a}  
\end{equation}
The number of solutions equals the number of coefficients, 
in accordance with the maximum-cut theorem. Therefore 
the four  coefficients appearing in Eq.~(\ref{Eq:DeltaN4a}) can be obtained by sampling the numerator 
on the four  solutions of  the eightfold cut, where  the decomposition~(\ref{Eq:N4a}) becomes 
\bea
{\cal N}^{(4,a)}_{1\cdots 8} =  \Delta_{12345678} \; .
\eea
 In our case we find that only
$\ci{12345678}{0}$ and $\ci{12345678}{1}$ are nonvanishing.

\paragraph{Sevenfold cut.}
The residue of the generic sevenfold cut $(i_1\cdots i_7)$ appearing in Eq.~(\ref{Eq:N4a}) 
can be parametrized using the results listed in  Table~\ref{Tab:Pentabox}.  For the process at hand, the structure of the numerator ensures that
residue can be parametrized just by  a constant term:
\bea
\Delta_{i_1\cdots i_7} = \ci{i_1\cdots i_7}{0} \; . 
\label{Eq:Delta7folda}
\eea
 The actual value of  $\ci{i_1\cdots i_7}{0}$ is obtained  by sampling the numerator and the
residue of the eightfold cut in correspondence of one solution  of the sevenfold cut, where
 \bea
 \Delta_{i_1\cdots i_7} = \frac{{\cal N}^{(4,a)}_{1\cdots 8} ( q, k )  -\Delta_{12345678}   }{\prod^8_{h \neq  i_1 \ldots i_7}D_{h}} \; .
 \label{Eq:Dec7folda}
 \eea

 The multipole decomposition of the integrand $\mathcal{I}^{(4,a)}_{1\cdots 8}$ becomes
 \bea
  {\cal I}^{(4,a)}_{1\cdots 8} ( q, k ) =  \frac{\ci{12345678}{0}  + \ci{12345678}{1} (q \cdot p_1) }{\db{1}\cdots \db{8} }
  +  \sum_{i=4}^{7}  \frac{ \ci{1\cdots (i-1)(i+1) \cdots 8}{0}}{\prod_{h\neq i}^8 \db{h}   } \, .
\label{eq:multipoledecplanarN4}
 \eea
This result also shows the decomposition of the integral as linear combination of two MIs with eight denominators and four MIs with seven denominators.

 \subsection{Crossed pentabox diagram}
 \label{SSec:NumPentacross}
\paragraph{Eightfold cut.}
The residue of the eightfold cut is parametrized  as (cf.  Table~\ref{Tab:Pentacross}) 
\begin{equation}
\Delta_{12345678} =  \ci{12345678}{0} +  \ci{12345678}{1} (q\cdot p_1) +
 \ci{12345678}{2} (k\cdot p_3) 
   +  \ci{12345678}{3} (k\cdot p_4)  \; .
 \label{Eq:DeltaN4b}  
\end{equation}
The coefficients are obtained sampling the numerator at the four solutions of the maximum cut 
$(12345678)$, where 
\bea
{\cal N}^{(4,b)}_{1\cdots 8} =  \Delta_{12345678} \; .
\eea
The only nonvanishing coefficients are $\ci{12345678}{0}$ and $\ci{12345678}{1}$.

\paragraph{Sevenfold cut.}
The generic  sevenfold cut  appearing in the multipole decomposition of $\mathcal{I}_{1\cdots 8}^{(4,b)}$ is
 $(i_1\cdots i_7) \in$   $\{ (1234568),$ $(1234578),$ $(1234678),$ $(1235678) \}$. The structure 
 of $\mathcal{N}^{(4,b)}_{1\cdots 8}$  guarantees that the only nonvanishing coefficient is the one of the 
 monomial $1$, i.e.
 \bea
\Delta_{i_1\cdots i_7} = \ci{i_1\cdots i_7}{0} \; . 
\eea
We sample the numerator and the residue of the eightfold cut at a solution  of the cut $(i_1\cdots i_7)$ and 
we get   $\ci{i_1\cdots i_7}{0}$  by using the relation
  \bea
 \Delta_{i_1\cdots i_7} = \frac{{\cal N}^{(4,a)}_{1\cdots 8} ( q, k )  -\Delta_{12345678}   }{\prod^8_{h \neq  i_1 \ldots i_7}D_{h}} \; .
 \eea
 Within the numerical  precision, the coefficients of the crossed pentabox  turn out  to be equal to 
 the ones of the planar pentabox.  This is expected since the two diagrams share the same numerator and also the denominators appearing in its decomposition are in common between the two.

The integrand $\mathcal{I}^{(4,b)}_{1\cdots 8}$ is decomposed as follows:
 \bea
  {\cal I}^{(4,b)}_{1\cdots 8} ( q, k ) =  \frac{\ci{12345678}{0}  + \ci{12345678}{1} (q \cdot p_1) }{\db{1}\cdots \db{8} }
  +  \sum_{i=4}^{7}  \frac{ \ci{1\cdots (i-1)(i+1) \cdots 8}{0}}{\prod_{h\neq i}^8 \db{h}   } \, .
  \label{Eq:MultiPb}
 \eea
As for the previous diagram, this result yields the decomposition of the integral as linear combination of two MIs with eight denominators and four MIs with seven denominators. 
 
 \subsection{Double pentagon diagram}
  \label{SSec:NumPentatwist}
\paragraph{Eightfold cut.}
  The parametrization of the  residue of the eightfold cut is given in Table~\ref{Tab:Pentatwist} and can be written as
\bea
\Delta_{12345678} &=&  \ci{12345678}{0}  +  
 \ci{12345678}{1}    (k\cdot p_3) + 
  \ci{12345678}{2}   (q\cdot p_5) \nn 
  &+&
   \ci{12345678}{3}   (q \cdot p_1)  +
    \ci{12345678}{4}    (q\cdot p_5)^2   
     +   \ci{12345678}{5}    (q\cdot p_5)(q \cdot p_1)  \nn
      &+&  \ci{12345678}{6}    (k\cdot p_3) (q\cdot p_5) +  \ci{12345678}{7}    (q\cdot p_5)^3 \, .
 \label{Eq:DeltaN4c}
\eea
The eightfold cut is a maximum cut, thus the eight solutions of the eightfold cut allow one to determine the coefficients in Eq.~(\ref{Eq:DeltaN4c}) using 
the relation
\bea
{\cal N}^{(4,b)}_{1\cdots 8} =  \Delta_{12345678} \; ,
\eea 
which holds at the solutions of the eightfold cut.   The nonvanishing coefficients
are  $\ci{12345678}{i} $  for $i \le 4$.

\paragraph{Sevenfold cut.}
The numerator has rank one and it is easy to see that 
the generic sevenfold residue $\Delta_{i_1\cdots i_7}$
entering Eq.~(\ref{Eq:N4c})  can be parametrized by a  constant term, i.e.
\bea
\Delta_{i_1\cdots i_7} = \ci{i_1\cdots i_7}{0} \; . 
\eea
The value of $\ci{i_1\cdots i_7}{0}$ is obtained by sampling the numerator and the residue of the eightfold cut 
at a solution of the cut $(i_1\cdots i_7)$, where the relation
  \bea
 \Delta_{i_1\cdots i_7} = \frac{{\cal N}^{(4,c)}_{1\cdots 8} ( q, k )  -\Delta_{12345678}   }{\prod^8_{h \neq  i_1 \ldots i_7}D_{h}} \; 
 \eea
holds.  The multipole decomposition of the integrand of the double pentagon reads as follows
\bea
 {\cal I}^{(4,c)}_{1\cdots 8} ( q, k ) &=&  
           \frac{\ci{12345678}{0}  + \ci{12345678}{1} (k \cdot p_3)}{\db{1}\cdots \db{8}} \nn
 &+&  \frac{ \ci{12345678}{2} (q \cdot p_5) + \ci{12345678}{3} (q \cdot p_1)}{\db{1}\cdots \db{8} } \nn
 &+&  \sum_{i=1}^{8}  \frac{ \ci{1\cdots (i-1)(i+1) \cdots 8}{0}}{\prod_{h \neq i}^8 \db{h}   } \; .
  \label{Eq:Multipole4C}
\eea
The corresponding decomposition of the integral is a linear combination of four MIs with eight denominators and eight MIs with seven denominators.

\subsection{Unitarity-based construction}

In the previous subsections as well as in the following sections, we apply 
the multiloop integrand-reduction method in the case of integrands  
provided by a diagrammatic 
representation of the scattering amplitude, where  the full dependence 
on the loop momenta is known.
The method can, however,  be applied also using a unitarity-based representation
of the integrands, where the latter is known  only in correspondence to multiple cuts 
in physical channels  as  a state sum over the
product of tree-level amplitudes. 

In this case,  the  integrand of the  (color ordered) amplitude in $\mathcal{N}=4$ SYM
reads as
\begin{equation}
\mathcal{I}^{(4)}(q,k)  = \sum_{ ( i_1\cdots i_8)} \frac{\Delta_{i_1\cdots i_8}}{D_{i_1}\cdots D_{i_8}}
+  \sum_{(i_1\cdots i_7)} \frac{\Delta_{i_1\cdots i_7}}{D_{i_1}\cdots D_{i_7}}  \, ,
\end{equation}
where the first (second) sum  runs over all the eightfold (sevenfold) cuts. 

The residue of the generic eightfold cut is given by
\begin{equation}
\Delta_{i_1\cdots i_8} = \mbox{Res}_{\, i_1\cdots i_8} \left \{  \mathcal{I}^{(4)} \right  \} \, ,
\end{equation}
where the (maximum-cut) residue $\mbox{Res}_{\, i_1\cdots i_8} \{  \mathcal{I}^{(4)} \} $
is the state sum over the product of seven three-point tree-level 
amplitudes.  

The residue of the generic sevenfold cut reads instead
\begin{equation}
\Delta_{i_1\cdots i_7} = \mbox{Res}_{\, i_1\cdots i_7}\left  \{  \mathcal{I}^{(4)}   - \sum_{ (i_1\cdots i_8 )  } 
 \frac{\Delta_{i_1\cdots i_8} }{D_{i_1}\cdots D_{i_8}} \right \} \, ,
\end{equation}
where $\mbox{Res}_{\, i_1\cdots i_7} \{  \mathcal{I}^{(4)} \}$  is the state-sum product of six  tree-level
amplitudes, while the sum  runs over all  the sets  $(i_1\cdots i_8)$  containing 
$(i_1\cdots i_7)$ as a subset.

The extension of the algorithm to lower cuts, if needed, is straightforward.

\section{Five-point amplitudes in  ${\cal N}=8$ SUGRA}
The  five-point amplitude in ${\cal N}=8$ SUGRA can be expressed in terms of 
the same six diagrams as in ${\cal N}=4$ SYM~\cite{Carrasco:2011mn}.  Again, 
the color ordered   amplitude   is given by a  sum over  the cyclic permutations of the external momenta. 
In  ${\cal N}=8$ SUGRA,   the  numerator of each integrand  is 
 obtained by squaring   the corresponding numerator   in ${\cal N}=4$ SYM, as 
 shown in~\cite{Carrasco:2011mn}.  
We apply the integrand reduction only to the three diagrams  depicted in 
Fig.~\ref{fig:diagram}, whose numerator  exhibits  a nontrivial dependence 
on the loop momenta.  In the following we  denote the integrand (numerator)  
of the diagram in Fig.~\ref{fig:diagram} $(i)$  by  $\mathcal{I}^{(8, i)}$   ( $\mathcal{N}^{(8, i)}$).

The numerators are of rank two in the loop momenta.  Following the same machinery as in the case of ${\cal N}=4$ SYM, we show that their decomposition can be expressed in terms of 8-, 7-, and 6-denominator integrands:
\bea
{\cal N}^{(8,x)}_{1\cdots 8} ( q, k ) &=&  \Delta_{12345678}  +  \sum_{i=1}^8  \Delta_{1\cdots (i-1)(i+1) \cdots 8} \db{i} \pagebreak[1]   \nn
&& +  \sum_{i<j =1} ^8  \Delta_{1\cdots (i-1)(i+1) \cdots (j-1)(j+1) \cdots 8} \db{i} \db{j} \,\,\, , \,\,\, x=a,b,c \; . 
\label{Eq:N8a}
\label{Eq:N8b}
\label{Eq:N8c}
\eea

The corresponding decomposition for the integrands ${\cal
  I}^{(4,a)}_{1\cdots 8}$, ${\cal I}^{(4,b)}_{1\cdots 8}$ and ${\cal
  I}^{(4,c)}_{1\cdots 8}$ reads
\bea
 {\cal I}^{(4,x)}_{1\cdots 8} ( q, k ) &=&  \frac{\Delta_{12345678}}{\db{1}\cdots \db{8} }  +  \sum_{i=1}^{8}  \frac{ \Delta_{1\cdots (i-1)(i+1) \cdots 8}}{\prod_{h \neq i}^8 \db{h}   } \nn
  && +  \sum_{i<j =1} ^8  \frac{\Delta_{1\cdots (i-1)(i+1) \cdots (j-1)(j+1) \cdots 8}}{\prod_{h \neq i,j}^8 \db{h}} \,\,\, , \qquad \,\,\, x=a,b,c \; . 
  \label{Eq:Multipole8}
\eea
Since the numerators $\mathcal{N}^{(8, a)}_{1\cdots 8}$ and
$\mathcal{N}^{(8, b)}_{1\cdots 8}$ are of rank two in $q$ and
independent of $k$, their decomposition is significantly simplified.
Indeed in these cases $\Delta_{1\cdots (i-1)(i+1)\cdots 8}=0$ for $i\neq
4,5,6,7$ and $\Delta_{1\cdots (i-1)(i+1) \cdots (j-1)(j+1) \cdots 8} =
0$ for $i,j\neq 4,5,6,7$.

\subsection{Seminumerical computation}
In this section we briefly describe the numerical  decomposition of the  three numerators. 
We checked the decomposition by verifying that at arbitrary values of $q$ and $k$
the numerator and its reconstruction are equal.

\subsubsection*{Planar pentabox}
\paragraph{Eightfold cut. }
The computation of the residue of the eightfold cut $\Delta_{12345678}$ follows the same pattern as the $\mathcal{N}=4$
SYM planar pentabox, described in  Section~\ref{SSec:NumPentabox}.  The nonvanishing coefficients are
$\ci{12345678}{0}$ and $\ci{12345678}{1}$. 

\paragraph{Sevenfold cut. }
The residue of the generic sevenfold cut $(i_1\cdots i_7)$ in Eq.~(\ref{Eq:N8a}) can be parametrized in terms 
of the monomials collected in Table~\ref{Tab:Pentabox}. The structure of the numerator guarantees that 
the residue contains  rank-one terms at most:
\bea
\Delta_{i_1\cdots i_7} &=& \ci{i_1\cdots i_7}{0} + \ci{i_1\cdots i_7}{1} (q+\mo_0)\cdot \mo_1 + 
\ci{i_1\cdots i_7}{2} (q+\mo_0)\cdot \mo_2 \nn 
&& + \ci{i_1\cdots i_7}{3} (k+\mo_3)\cdot \mo_4 + \ci{i_1\cdots i_7}{4} (k+\mo_3)\cdot \mo_5 \, .
\label{Eq:7foldaPAR}
\eea
The momenta $\mo^\mu_i$ depends on the cut $(i_1\cdots i_7)$.  The actual value of 
the coefficients can be obtained by sampling on five independent solutions of the sevenfold cut,
where 
\bea
 \Delta_{i_1\cdots i_7} = \frac{{\cal N}^{(8,a)}_{1\cdots 8} ( q, k )- \Delta_{12345678}}{\prod^8_{h \neq i_1,\ldots , i_7}D_{h}}     \; . 
\eea
In this case, all the coefficients but $\ci{i_1\cdots i_7}{0}$, $\ci{i_1\cdots i_7}{1}$, 
and $\ci{i_1\cdots i_7}{2}$  vanish.

\paragraph{Sixfold cut. }
The numerator ${\cal N}^{(8,a)}_{1\cdots 8}$ has rank two in $q$ and
is independent of $k$.  The only nonvanishing term in the residue of
the generic sixfold cut $(i_1\cdots i_6)$ in~(\ref{Eq:N8a}) is the constant; therefore we have
\bea
\Delta_{i_1\cdots i_6} &=& \ci{i_1\cdots i_6}{0} \, .
\label{Eq:Para6N8}
\eea
The actual value of the constant can be obtained evaluating the decomposition~(\ref{Eq:N8a}) at one solution of the sixfold cut, where
\bea
 \Delta_{i_1\cdots i_6} = \frac{{\cal N}^{(8,a)}_{1\cdots 8} ( q, k )  - \Delta_{12345678} }{\prod^8_{h \neq i_1,\ldots , i_6}D_{h}} 
 -    \sum_{ h=4   \atop  h \neq i_1,\ldots i_6  ,
%
  }^7   
  \frac{\Delta_{i_1 \cdots  h \cdots i_6}   }{D_{h}}
    \; . 
 \label{Eq:Dec6folda}
\eea

After polynomial fitting of $\Delta_{12345678}$, $\Delta_{i_1 \cdots
  i_7}$ and $\Delta_{i_1 \cdots i_6}$, the resulting multipole
decomposition of Eq.~\eqref{Eq:Multipole8} contains 20 nonvanishing
coefficients two of which are spurious (i.e.\ their contribution
vanishes upon integration) while the others give rise to MIs, namely
two with eight denominators, ten with seven denominators and six with
six denominators.

\subsubsection*{Crossed pentabox}
As in the $\mathcal{N}=4$ SYM  case,   
the crossed pentabox in Fig~\ref{fig:diagram} $(b)$ has the same numerator and the same decomposition as the planar pentabox. Therefore
the coefficients of the former  are exactly the same as the coefficients of the latter.

\subsubsection*{Double pentagon}
\paragraph{Eightfold cut. }
The computation  residue  of the eightfold cut of the double pentagon  follows the same lines of the $\mathcal{N}=4$
SYM double  pentagon  (see  Section~\ref{SSec:NumPentatwist}).   The parametrization of the residue is given in Eq.~(\ref{Eq:DeltaN4c}). In this 
case the  only vanishing coefficient is $\ci{12345678}{7}$.

\paragraph{Sevenfold cut. }
The  residue  $\Delta_{i_1\cdots i_7}$ of the generic cut $(i_1\cdots i_7)$   can be parametrized using Eq.~(\ref{Eq:7foldaPAR}). 
At the sevenfold cut $(i_1\cdots i_7)$  the decomposition~(\ref{Eq:N8c}) reduces to
\bea
 \Delta_{i_1\cdots i_7} = \frac{{\cal N}^{(8,c)}_{1\cdots 8} ( q, k )- \Delta_{12345678}}{\prod^8_{h \neq i_1,\ldots , i_7}D_{h}}     \; . 
 \label{Eq:Dec7fold}
 \eea
The coefficients are then  computed by sampling Eq.~(\ref{Eq:Dec7fold}) at  five solutions of the sevenfold cut.  The nonvanishing ones are those multiplying constant or linear terms in the loop momenta.

\paragraph{Sixfold cut. }
The residue $\Delta_{i_1\cdots i_6}$ of the generic sixfold cut $(i_1\cdots i_6)$ can be parametrized by a constant, 
as in Eq.~(\ref{Eq:Para6N8}).  The constant is computed  using one solution of the sixfold cut and the expression
of the decomposition~(\ref{Eq:N8c}) at the sixfold cut:
\bea
\Delta_{i_1\cdots i_6} = \frac{{\cal N}^{(8,a)}_{1\cdots 8} ( q, k )-\Delta_{12345678} }{\prod^8_{h \neq i_1,\ldots , i_6}D_{h}} 
  -    \sum_{h \neq i_1,\ldots ,i_6}^8   \frac{\Delta_{i_1 \cdots  h \cdots i_6}   }{D_{h}}
    \; . 
 \label{Eq:Dec6foldc}
\eea
We find that $\Delta_{123456}=0$, while the residues of all the other sixfold cuts are nonvanishing.

After polynomial fitting of $\Delta_{12345678}$, $\Delta_{i_1 \cdots
  i_7}$ and $\Delta_{i_1 \cdots i_6}$, the resulting multipole
decomposition of Eq.~\eqref{Eq:Multipole8} contains in this case 74
nonvanishing coefficients, four of which are spurious.  The integral
can be decomposed as a linear combination of seven MIs with eight
denominators, 36 MIs with seven denominators and 27 MIs with six
denominators.

\def\Nhat{\mathcal{R}}
\def\Ntil{\mathcal{T}}

\section{Analytic integrand reduction}
\label{Sec:analytic}
In this section we perform the reduction of the five-point diagrams 
analytically.  We apply a two-loop generalization of the integrand
reduction through Laurent expansion formulated
in~\cite{Mastrolia:2012bu}.  As in the one-loop case, the Laurent
expansion allows one to find simpler formulas for the coefficients
entering the decomposition. Moreover, the subtraction of the higher
residues can be performed at the coefficient level rather than at the
integrand level. Indeed the Laurent expansion makes each function
entering the reduction separately polynomial. Therefore the
subtraction can be omitted during the reduction and accounted for
correcting the reconstructed coefficients.  For simplicity we will
focus on the rank-one numerators in the five-point integrands  of ${\cal N} =
4$ SYM. The method can, however,  be extended to higher-rank 
numerators as described in Section \ref{sec:higher-rank} for the 
planar pentabox diagram in $\N=8$ SUGRA.

%

\subsection{Planar pentabox diagram}
\label{sec:analytic-pentabox}
The analytic decomposition of the pentabox diagram in Fig.\
\ref{fig:diagram} $(a)$ we are about to discuss, is valid for any numerator of the
type~(\ref{Eq:pentaboxN4}), i.e. for any rank-one numerator depending
on $q$ only.  Indeed our computation is carried out for generic
$\vi^\mu$ and $\beta_1$; the results for $\N = 4$ SYM will be
recovered at the very end, using Eqs.\ \eqref{eq:valphan4pbv} and
\eqref{eq:valphan4pbalpha}.

\paragraph{Eightfold cut.}
The four solutions of the eightfold cut  $(12345678)$  are
  \begin{align}
   \left (q^\mu_1,k^\mu_1\right ) = \left ( \frac{\spa{4\, 5}}{\spa{3\, 5}}\frac{\spab{3|\gamma^\mu|4}}{2}, \;  \frac{\spa{3\, 2}}{\spa{1\, 3}}\frac{\spab{1|\gamma^\mu|2}}{2} \right ),  \qquad &
    \left (q^\mu_2,k^\mu_2\right ) = \left (\frac{\spa{4\, 5}}{\spa{3\, 5}}\frac{\spab{3|\gamma^\mu|4}}{2}, \;  \frac{\spa{1\, 5}}{\spa{2\, 5}}\frac{\spab{2|\gamma^\mu|1}}{2} \right ), \nn
    \left (q^\mu_3,k^\mu_3\right ) =  \left (  \frac{\spb{4\, 5}}{\spb{3\, 5}}\frac{\spab{4|\gamma^\mu|3}}{2},\;  \frac{\spb{1\, 5}}{\spb{2\, 5}}\frac{\spab{1|\gamma^\mu|2}}{2} \right ), \qquad   & 
    \left (q^\mu_4,k^\mu_4\right )  = \left ( \frac{\spb{4\, 5}}{\spb{3\, 5}}\frac{\spab{4|\gamma^\mu|3}}{2},\;  \frac{\spb{3\, 2}}{\spb{1\, 3}}\frac{\spab{2|\gamma^\mu|1}}{2} \right ). \nonumber    
  \end{align}
The general parametrization of  the residue $\Delta_{12345678}$ is given  in Eq.~(\ref{Eq:DeltaN4a}).
The simple form of the numerator $\mathcal{N}^{(4,a)}_{1\cdots 8}$ implies that 
the coefficients $\ci{12345678}{2}$ and $\ci{12345678}{3}$  
vanish. The nonvanishing coefficients  are obtained by sampling at the  solutions $(q_1,k_1)$ and $(q_3,k_3)$ only. The outcome is
\bea
\ci{12345678}{0} &=&-\frac{1}{ \spa{5 \, 4 } \spa{3 \, 1 } \spb{5 \, 3 } \spb{4 \, 1 }- \spa{5 \, 3 } \spa{4 \, 1 } \spb{5 \, 4 } \spb{3 \, 1 }} \nn
  &&  \quad \times \Big(   \spa{5 \, 4 } \spa{4 \, 1 } \langle3 |\vi | 4] \spb{5 \, 4 } \spb{3 \, 1 }-  \spa{5 \, 4 } \spa{3 \, 1 }  \langle 4 |\vi | 3] \spb{5 \, 4 } \spb{4 \, 1 } \nn
  &&   \qquad - \beta_1 \spa{5 \, 4 } \spa{3 \, 1 } \spb{5 \, 3 } \spb{4 \, 1 } + \beta_1 \spa{5 \, 3 } \spa{4 \, 1 } \spb{5 \, 4 } \spb{3 \, 1 } \Big) \\
\ci{12345678}{1}&=& -2 \frac{ \spa{5 \, 4 }  \langle3 |\vi | 4] \spb{5 \, 3 }- \spa{5 \, 3 }  \langle 4  |\vi | 3]  \spb{5 \, 4 }}{ \spa{5 \, 4 } \spa{3 \, 1 } \spb{5 \, 3 } \spb{4 \, 1 }- \spa{5 \, 3 } \spa{4 \, 1 } \spb{5 \, 4 } \spb{3 \, 1 }}.
\eea

\paragraph{Sevenfold cuts.}
We discuss the generic sevenfold cut $(i_1\cdots i_7)$ appearing the decomposition~(\ref{Eq:N4a}). For later convenience, we define
the uncut propagator 
\bea
D_{i_8}(q,k) = (q+P_{i_8})^2, \quad \mbox{with} \; i_8 \in \{1,\ldots,8\}  \quad  \mbox{and} \quad  i_8 \neq i_1,\cdots , i_7.
\label{Eq:Di8}
\eea
The momentum $P^\mu_{i_8}$ is a linear combination of external momenta and it  can be inferred from Fig.~\ref{fig:diagram} $(a)$, for instance
$P^\mu_7 = -p^\mu_4-p^\mu_5$.
The simplicity of the numerator allows one to parametrize 
 the residue using  the constant term  $\ci{i_1\cdots i_7}{0}$; cf. Eq.~(\ref{Eq:Delta7folda}). 
 
Every sevenfold cut of this topology exhibits a $t$-dependent 
family of solutions of the type
  \bea
  \left (q^\mu_1, k^\mu_1 \right )  & =&\left (  v^\mu_{q_1, 1}\, t + v^\mu_{q_1, 0},  \;  v^\mu_{k_1}  \right ) \; . 
  \label{Eq:PentaboxSolG}
 \eea
The coefficient  $\ci{i_1\cdot i_7}{0}$ can be computed evaluating Eq.~(\ref{Eq:Dec7folda})
 at the solutions~(\ref{Eq:PentaboxSolG}). 
 The computation can be simplified performing a Laurent expansion around  $t = \infty$:
 \bea
\left [   
\frac{\mathcal{N}^{(4,a)}_{1\cdots 8}(q_1, k_1) }{  D_{i_8}(q_1,k_1)  }
-  \frac{ \Delta_{12345678}(q_1,k_1)  }{  D_{i_8}(q_1,k_1)  } 
 \right ]_{t\to \infty} = 
\ci{i_1\cdots i_7}{0}
 \eea
 Indeed in general  
%
%
neither 
\bea
 \frac{  \mathcal{N}^{(4,a)}_{1\cdots 8}(q_1, k_1)  }{  D_{i_8}(q_1,k_1)  }   \qquad \mbox{nor} \qquad \frac{ \Delta_{12345678}(q_1,k_1)  }{  D_{i_8}(q_1,k_1)  } 
\eea
are polynomial in $t$ but  only their difference is.  However, their truncated Laurent expansion obtained neglecting $\O(1/t)$ terms 
is polynomial in $t$, namely in this case a constant
{\small \bea
\left . \frac{  \mathcal{N}^{(4,a)}_{1\cdots 8}(q_1, k_1) }{  D_{i_8}(q_1,k_1)  } \right  |_{t\to \infty} =  \ni{i_1\cdots i_7}{0} + \mathcal{O}\left ( \frac{1}{t}  \right ),  \qquad
\left .  \frac{ \Delta_{12345678}(q_1,k_1)  }{  D_{i_8}(q_1,k_1)  } \right  |_{t\to \infty} = \bi{i_1\cdots i_7}{0} + \mathcal{O}\left ( \frac{1}{t}  \right ) . \qquad 
\label{Eq:Exp4a}
\eea }
Therefore, similarly to the one-loop case~\cite{Mastrolia:2012bu}, the coefficients
$\ni{i_1\cdots i_7}{0}$ and $\bi{i_1\cdots i_7}{0}$ can be computed separately 
obtaining the coefficient $\ci{i_1\cdots i_7}{0}$ by their difference:
\bea
\ci{i_1\cdots i_7}{0} = \ni{i_1\cdots i_7}{0} - \bi{i_1\cdots i_7}{0}\; .
\eea 
The subtraction can be performed at the coefficient level rather than at the 
integrand level. Moreover the known structure of  $\Delta_{12345678}$ allows one  to compute 
the coefficient $\bi{i_1\cdots i_7}{0}$  once and for all, irrespective of the actual form of the 
numerator:
\bea
\bi{i_1\cdots i_7}{0} =  \frac{ \ci{12345678}{1}  (p_1\cdot v_{q_1,1})  }{ 2 (P_{i_8}+v_{q_1,0}) \cdot v_{q_1,1}  } \; .
\eea
The coefficients $\ni{i_1\cdots i_7}{0}$  read as follows:
\bea
\ni{1234568}{0} = -\frac{\vi \cdot  v_{q_1,1}}{ p_5 \cdot  v_{q_1,1}}  \qquad &\mbox{with}& \qquad   v^\mu_{q_1,1} = \frac{\langle 3 |\gamma^\mu | 4]   }{2} , \nn
\ni{1234678}{0} = \frac{\vi \cdot  v_{q_1,1}}{  p_3 \cdot  v_{q_1,1}}  \qquad  &\mbox{with}& \qquad   v^\mu_{q_1,1} = \frac{\langle 5 |\gamma^\mu | 4]   }{2}, \nn
\ni{1234578}{0} = -\frac{\vi \cdot  v_{q_1,1}}{ p_4 \cdot  v_{q_1,1}}  \qquad  &\mbox{with}& \qquad   v^\mu_{q_1,1} = \frac{\langle 3 |\gamma^\mu | P_{345}]   }{2}, \nn
\ni{1235678}{0} = -\frac{\vi \cdot  v_{q_1,1}}{ p_4 \cdot  v_{q_1,1}}  \qquad  &\mbox{with}& \qquad   v^\mu_{q_1,1} = \frac{\langle 5 |\gamma^\mu | P_{543}]   }{2}.
\eea
The massless momentum $P_{abc}$ is defined as 
\bea
P_{abc}^\mu \equiv p_b^\mu + p_c^\mu  - \frac{s_{bc}}{2(p_b+p_c)\cdot p_a} p_a^\mu \; .
\eea
%
%

As already stated, the foregoing discussion applies to any numerator
of the form given in Eq.~\eqref{Eq:pentaboxN4}.  By using the explicit
expressions of $\vi$ and $\beta_1$ given in
Eqs.~\eqref{eq:valphan4pbv} and \eqref{eq:valphan4pbalpha}, we can write
down the results for the coefficients in $\N=4$ SYM in terms of the
functions $\gamma$ defined in Eq~\eqref{eq:carrascogamma12345}
\begin{align}
\label{eq:coeffsan4}
  \ci{12345678}{0} = {} &  \frac{1}{2}  \left ( \gamma_{12 345} (s_{23}-s_{13}-s_{45}) + s_{12} (\gamma_{34 125}+\gamma_{35  124}+\gamma_{45  123}  ) \right ) \nn
  \ci{12345678}{1} = {} &  -2\, \gamma_{12 345} \nn
  \ci{1234568}{0} = {} & \frac{1}{4}(-\gamma_{35124}-\gamma_{45123}+2\, \gamma_{12345}) \nn
  \ci{1234578}{0} = {} & \frac{1}{4}(-\gamma_{34125}+\gamma_{35124}+2\, \gamma_{45123}) \nn
  \ci{1234678}{0} = {} & \frac{1}{4}(-\gamma_{35124}-\gamma_{34125}-2\, \gamma_{12345}) \nn
  \ci{1235678}{0} = {} & \frac{1}{4}(-\gamma_{45123}+\gamma_{35124}+2\, \gamma_{34125}).
\end{align}

The complete integrand decomposition is obtained plugging the
coefficients of Eq.~\eqref{eq:coeffsan4} in
Eq.~\eqref{eq:multipoledecplanarN4}.  These results are in agreement
with the ones found in the numerical computation of
Section~\ref{SSec:NumPentabox}.

\subsection{Crossed pentabox diagram}
As already noticed in Section~\ref{SSec:NumPentacross},
the crossed pentabox of  Fig.\ \ref{fig:diagram} $(b)$ 
and the planar pentabox have the same decomposition.
Indeed the numerator  $\mathcal{N}^{(4,a)}_{1\cdots 8}$  and the numerator
$\mathcal{N}^{(4,b)}_{1\cdots 8}$,  Eq.~(\ref{Eq:pentacrossN4}), are equal and are decomposed in terms of the 
same denominators; cf. Eqs.~(\ref{Eq:MultiPb}) and~(\ref{eq:multipoledecplanarN4}) multiplied by $D_{i_1}\ldots D_{i_8}$.
Therefore the  coefficients appearing in  the multipole  decomposition~(\ref{Eq:MultiPb}) 
are equal to the corresponding ones appearing in the decomposition~(\ref{eq:multipoledecplanarN4}) of the planar pentabox.

\subsection{Double  pentagon diagram}
\label{sec:analytic-pentatwist}
The numerator $\mathcal{N}^{(4,c)}_{1\cdots 8}$, Eq.~(\ref{eq:5pwn4symnumpar}), 
is highly symmetric in  $q$ and $k$. Indeed the dependence on $q$ and $k$ can be disentangled and 
  $\mathcal{N}^{(4,c)}_{1\cdots 8}$  can be cast in the following form:
\bea
\mathcal{N}^{(4,c)}_{1\cdots 8}(q,k) =  \Nhat^{(4,c)}_{1\cdots 8}(q)- \Ntil^{(4,c)}_{1\cdots 8}(k) \, ,
\label{eq:n4ceqNhat}
\eea
where 
\bea
 \Nhat^{(4,c)}_{1\cdots 8} (\ell) \equiv 2 \ell \cdot \wu + \beta_2,  \qquad  \Ntil^{(4,c)}_{1\cdots 8} (k)  =  \Nhat^{(4,c)}_{1\cdots 8} (k) \, |_{p_1 \leftrightarrow p_3,\, p_4 \leftrightarrow p_5} \; 
\label{eq:n4chat}
\eea
The vector  $\wu$ and the constant $\beta_2$ are defined in Eqs.~(\ref{eq:valphan4ptv}) and~(\ref{Eq:beta2}), respectively. 
Although in principle there are no additional issues in performing the reduction of the full numerator, the computation can be 
 simplified by performing the reduction of
 $\Nhat^{(4,c)}_{1\cdots 8}(q)$ only. The reduction
 of   $\Ntil^{(4,c)}_{1\cdots 8}(q)$, and thus  of the full numerator $\mathcal{N}^{(4,c)}_{1\cdots 8}(q,k)$, 
is then obtained by means of the substitutions~(\ref{Eq:transform}).
The  
numerator $\Nhat^{(4,c)}_{1\cdots 8} (q)$ depends on $q$ only and  can be decomposed as follows:
\bea
\Nhat^{(4,c)}_{1\cdots 8} (q) =  \Delta_{12345678}  +  \sum_{i=1}^3  \Delta_{1\cdots (i-1)(i+1) \cdots 8} \db{i} \; .
\label{Eq:NhatDec}
\eea


\paragraph{Eightfold cut.} The solutions of the eightfold cut  $(12345678)$ are eight and 
they can be used to compute the eight coefficients parametrizing the residue 
\bea
\Delta_{12345678} &=&  \hci{12345678}{0}  +  
 \hci{12345678}{1}    (k\cdot p_3) + 
 \hci{12345678}{2}   (q\cdot p_5) \nn 
  &+&
   \hci{12345678}{3}   (q \cdot p_1)  +
    \hci{12345678}{4}    (q\cdot p_5)^2   
     +   \hci{12345678}{5}    (q\cdot p_5)(q \cdot p_1)  \nn
      &+&  \hci{12345678}{6}    (k\cdot p_3) (q\cdot p_5) +  \hci{12345678}{7}    (q\cdot p_5)^3 \; .
 \label{Eq:DeltaNhat4c}
\eea
The rank of $\Nhat^{(4,c)}_{1\cdots 8}$ implies that 
for $i\ge 4$ $\hci{12345678}{i} =0$. The simplicity of the numerator 
 simplifies the computation  even further. Indeed decomposing  $\wu^\mu$ in 
 the basis $\{p_i\}_{i=1,\cdots,4}$, and using the conditions $D_1=D_2=D_3=0$
 we get
 \bea
\hci{12345678}{0} =\beta_2,  \qquad  \hci{12345678}{1} =0  \; .
 \eea
 The two remaining coefficients can be obtained  by sampling the numerator 
 on two solutions of the eightfold cut, e.g.
\bea
  (k_1^\mu, q_1^\mu) &=& \left ( \frac{\langle 4 \,  1 \rangle}{\langle4 \, 5 \rangle}
 \frac{    \langle 5 |\gamma^\mu |1 ] }{2} , \;  \frac{\spa{3 \, 5 }}{\spa{4 \, 5 }}\,  \frac{    \langle 4 |\gamma^\mu |3 ] }{2} \right ) \, , \nn
   (k_2^\mu, q_2^\mu)  &=& \left (\frac{[ 4 \,  1 ]}{ [ 4 \, 5 ]} \, 
    \frac{    \langle 1 |\gamma^\mu |5 ] }{2} , \;  \frac{\spb{3 \, 5 }}{\spb{4 \, 5 }}   \frac{    \langle 3 |\gamma^\mu |4 ] }{2} \right ) \, .\nonumber 
%
%
\eea
The missing coefficients read as follows:
\begin{align}
  &  \hci{12345678}{2} = 
-2\frac{  \langle 4 |1 |3 ]  \langle 3 |\wu |4 ]  -  \langle 3 |1 |4 ]  \langle 4 |\wu |3 ] }{  \langle 4 |5 |3 ]\langle 3 |1 |4 ]  -     \langle 3 |5 |4 ]\langle 4 |1 |3 ]  }\, , \nn
  & \hci{12345678}{3} =   2\frac{  \langle 4 |5 |3 ]  \langle 3 |\wu|4 ]  -  \langle 3 |5 |4 ]  \langle 4 |\wu|3 ] }{  \langle 4 |5 |3 ]\langle 3 |1 |4 ]  -     \langle 3 |5 |4 ]\langle 4 |1 |3 ]  } . 
\end{align}
%
%
%
\paragraph{Sevenfold cuts.}  We consider the generic sevenfold cut $(i_1\cdots i_7)$ appearing in the decomposition~(\ref{Eq:NhatDec}).
 The solutions of the cut can be cast into  one-parameter families. In particular each cut allows for a solution with the following asymptotic behavior
\bea
(k_1^\mu, \; q_1^\mu)  = \left ( v_{k_1,1}^\mu  t +  v_{k_1,0}^\mu  + \mathcal{O}\left (\frac{1}{t} \right ), \;v_{q_1,1}^\mu  t  +   v_{q_1,0}^\mu  + \mathcal{O}\left ( \frac{1}{t} \right )   \right )  \qquad \mbox{for} \; t\to \infty \, .
\eea 
 We compute the coefficient $\hci{1345678}{0}$  by evaluating the 
decomposition~(\ref{Eq:NhatDec}) at one solution of the sevenfold cut,
and by expanding around $t=\infty$:
\bea
\left [   
\frac{\Nhat^{(4,c)}_{1\cdots 8}(q_1) }{  D_{i_8}(q_1,k_1)  }
-  \frac{ \Delta_{12345678}(q_1,k_1)  }{  D_{i_8}(q_1,k_1)  } 
 \right ]_{t\to \infty} = 
\hci{i_1\cdots i_7}{0}  \; .
\eea
 The denominator $D_{i_8}$ is written in terms of $P_{i_8}^\mu$ as in Eq.~(\ref{Eq:Di8}). The actual form
 of $P_{i_8}^\mu$ is inferred from Fig.~\ref{fig:diagram} $(c)$.  Also in this case the $t \to \infty$ limit makes
 both 
 \bea
 \frac{\Nhat^{(4,c)}_{1\cdots 8}(q_1) }{  D_{i_8}(q_1,k_1)  } \qquad \mbox{and} \qquad
  \frac{ \Delta_{12345678}(q_1,k_1)  }{  D_{i_8}(q_1,k_1)  } 
 \eea
polynomial in $t$,  
 {\small \bea
\left .  \frac{\Nhat^{(4,c)}_{1\cdots 8}(q_1) }{  D_{i_8}(q_1,k_1)  } \right |_{t \to \infty}= \hni{i_1\cdots i_7}{0}  + \mathcal{O}\left ( \frac{1}{t}  \right ), \qquad 
\left .  \frac{ \Delta_{12345678}(q_1,k_1)  }{  D_{i_8}(q_1,k_1)  }  \right |_{t \to \infty}= \hbi{i_1\cdots i_7}{0}  + \mathcal{O}\left ( \frac{1}{t}  \right ). 
 \eea}
 The coefficients  $\hni{i_1\cdots i_7}{0} $ and $\hbi{i_1\cdots i_7}{0} $ can be 
 computed separately, and
 \bea
\hci{i_1\cdots i_7}{0} =  \hni{i_1\cdots i_7}{0} - \hbi{i_1\cdots i_7}{0} \; .
 \eea
 Therefore the subtraction can be performed at the coefficient level via the universal function 
  \bea
 \hbi{i_1\cdots i_7}{0}  =  \frac{    \hci{12345678}{2}  ( v_{q_1,1} \cdot p_5)  +  \hci{12345678}{3}  ( v_{q_1,1} \cdot p_1)   }{2 (v_{q_1,0} + P_{i_8}) \cdot v_{q_1,1}}   \; . 
 \eea
 The  coefficients $\hni{i_1\cdots i_7}{0} $ are given by
 \bea
\hni{1345678}{0} =  - \frac{\wu \cdot v_{q_1} }{p_3 \cdot v_{q_1}}
  ,  \quad &\mbox{with}& \quad   v^\mu_{q_1} = \ETA_1 \frac{\langle 4 | \gamma^\mu | 3]}{2} + \ETA_2 p_4^\mu, \nn
\hni{1245678}{0} =  \frac{\wu \cdot v_{q_1} }{p_4 \cdot v_{q_1}}  ,
 \quad  &\mbox{with}& \quad   v^\mu_{q_1} = \ETA_3 \frac{\langle 4 | \gamma^\mu | 3]}{2} + \ETA_4 p_3^\mu, \nn
\hni{1234578}{0} = \frac{\wu \cdot v_{q_1} }{p_3 \cdot v_{q_1} }    ,   \quad  &\mbox{with}& \quad   v^\mu_{q_1} =
\ETA_5 (p_3^\mu-p_4^\mu) + \ETA_6 \frac{\langle3 |\gamma^\mu| 4]}{2} + \ETA_7 \frac{\langle4 |\gamma^\mu| 3]}{2}  .
\eea
We define
\bea
\ETA_1 &=& -\frac{\spa{5 \, 2 } \spb{4 \, 1 }}{ \spa{4 \, 2 } \spb{3 \, 4 } } \, , \nn 
\ETA_2 &=& \frac{ \spa{5 \, 2 } \spb{3 \, 1 }}{ \spa{4 \, 2 } \spb{3 \, 4 } } \, , \nn 
\ETA_3 &=& \frac{\spa{3 \, 5 } \spb{1 \, 2 } }{ \spa{3 \, 4 } \spb{3 \, 2 } } \, , \nn 
\ETA_4 &=& -\frac{ \spa{4 \, 5 } \spb{1 \, 2 }}{ \spa{3 \, 4 } \spb{3 \, 2 } } \, , \nn 
\ETA_5 &=&  \frac{ -\SIG_1 + \sqrt{\SIG_1^2 - 4 \SIG_2 \SIG_3} }{2 \SIG_2} \, , \nn 
\ETA_6 &=& \frac{2 \SIG_2 \SIG_8-\SIG_1 \SIG_4 +  \SIG_4 \sqrt{\SIG_1^2 - 4 \SIG_2 \SIG_3}}{2 \SIG_2 \SIG_5} \, , \nn 
\ETA_7 &=& -\frac{2  \SIG_2 \SIG_7 - \SIG_1 \SIG_6  + \SIG_6 \sqrt{\SIG_1^2 - 4 \SIG_2 \SIG_3} }{ 2 \SIG_2 \SIG_5} \, ,
\eea
in terms of
\begin{align}
\SIG_1 = {} & - 4 (\SIG_7 \SIG_4 + \SIG_6 \SIG_8)   (p_1 \cdot p_2)  \, , \nn 
\SIG_2 = {} & - 4 (\SIG_6 \SIG_4 - \SIG_5^2)   (p_1 \cdot p_2)  \, , \nn 
\SIG_3 = {} & - 4 \SIG_7 \SIG_8  (p_1 \cdot p_2)  \, , \nn 
\SIG_4 = {} & 2  (p_1 \cdot p_5)  \spa{4 \, 5 } \spb{3 \, 1 } - 2  (p_2 \cdot p_5)  \spa{4 \, 5 } \spb{3 \, 1 } + 
	  \spa{3 \, 5 } \spa{4 \, 2 } \spb{3 \, 1 } \spb{3 \, 2 } - 
	  \spa{4 \, 5 } \spa{4 \, 2 } \spb{3 \, 2 } \spb{4 \, 1 } \, , \nn 
\SIG_5 = {} & -\spa{3 \, 5 } \spa{4 \, 2 } \spb{3 \, 2 } \spb{4 \, 1 } + 
	  \spa{3 \, 2 } \spa{4 \, 5 } \spb{3 \, 1 } \spb{4 \, 2 } \, , \nn 
\SIG_6 = {} & 2 (p_2,\cdot p_3)\spa{3 \, 5 } \spb{4 \, 1 }
         -2  (p_2 \cdot p_5)  \spa{3 \, 5 } \spb{4 \, 1 }+
	    \spa{3 \, 5 } \spa{3 \, 2 } \spb{3 \, 1 } \spb{4 \, 2 } - 
	    \spa{3 \, 2 } \spa{4 \, 5 } \spb{4 \, 1 } \spb{4 \, 2 } \, , \nn 
\SIG_7 = {} & \spa{3 \, 5 } \spa{5 \, 2 } \spb{4 \, 1 } \spb{1 \, 2 } \, , \nn 
\SIG_8 = {} & \spa{4 \, 5 } \spa{5 \, 2 } \spb{3 \, 1 } \spb{1 \, 2 }.
\end{align}

This completes the reduction of a generic numerator
$\Nhat^{(4,c)}_{1\cdots 8}$ of the form given in
Eq.~\eqref{eq:n4chat}.  As previously stated, the reduction of the
full numerator $\mathcal{N}^{(4,c)}_{1\cdots 8}$ of $\N=4$ SYM can be
recovered from the one discussed in this section, by means of Eqs. \eqref{eq:n4ceqNhat} and \eqref{eq:n4chat}.  To that purpose, we
observe that the substitutions~(\ref{Eq:transform}) give the
one-to-one mapping between denominators
\begin{equation*}
    D_1\leftrightarrow D_4, \qquad D_2\leftrightarrow D_6, \qquad D_3\leftrightarrow D_5, \qquad D_7\leftrightarrow D_8.
\end{equation*}
Putting everything together and using the definitions of $\wu$ and
$\beta_2$ of Eqs.~(\ref{eq:valphan4ptv}) and~(\ref{Eq:beta2}) we
recover the full multipole decomposition of
Eq.~\eqref{Eq:Multipole4C}, with coefficients
  \begin{align}
    \ci{12345678}{0} & = \frac{1}{4} \Big( \gamma_{34125} ( 2
    s_{13} - 2 s_{35} + 2 s_{23} + s_{15} + s_{34} ) \nn & \qquad
    -\gamma_{15234} ( 2 s_{13} - 2 s_{35} + 2 s_{12} + s_{15} + s_{34} )
    \nn & \qquad - (\gamma_{23145}+\gamma_{24135})\, s_{23} - (\gamma_{25134}-\gamma_{12345}) (  s_{35} -s_{14} - s_{12} )
    \Big)\, , \displaybreak[0] \nn
    \ci{12345678}{1} & = -2\, \gamma_{34125} \, ,  \nn
    \ci{12345678}{2} & = \frac{1}{2}( 2 \gamma_{34125} 
    -\gamma_{23145} -\gamma_{24135} - 2\gamma_{15234} 
    +\gamma_{25134} - \gamma_{12345})\, ,  \displaybreak[0] \nn
    \ci{12345678}{3} & = \frac{1}{2}( 2 \gamma_{34125} 
    -\gamma_{23145} -\gamma_{24135} + 2\gamma_{15234} 
    +\gamma_{25134} - \gamma_{12345})\, ,  \displaybreak[0] \nn
    \ci{2345678}{0} & = \frac{1}{4}(  2 \gamma_{34125} -\gamma_{23145} +\gamma_{24135} ) \, , \displaybreak[0]  \nn
    \ci{1345678}{0} & = \frac{1}{4}( - 3 \gamma_{34125} + 2\gamma_{23145} +\gamma_{24135} -\gamma_{25134} + 
      \gamma_{12345} )\, ,  \displaybreak[0] \nn
    \ci{1245678}{0} & = \frac{1}{4}( \gamma_{34125} -\gamma_{23145} - 2\gamma_{24135} +\gamma_{25134} - 
      \gamma_{12345} )\, ,  \displaybreak[0] \nn
    \ci{1235678}{0} & = \frac{1}{4}( - 2\gamma_{15234} -\gamma_{25134} - \gamma_{12345} ) \, , \displaybreak[0] \nn
    \ci{1234678}{0} & = \frac{1}{4}(- 2 \gamma_{34125} +\gamma_{15234} +\gamma_{25134} )\, ,  \displaybreak[0] \nn
    \ci{1234578}{0} & = \frac{1}{4}( 2 \gamma_{34125} +\gamma_{15234} + \gamma_{12345} )\, ,  \displaybreak[0] \nn
    \ci{1234568}{0} & = \frac{1}{4}( 2 \gamma_{34125} +\gamma_{25134} - \gamma_{12345} )\, ,  \displaybreak[0] \nn
    \ci{1234567}{0} & = \frac{1}{4}(- 2 \gamma_{34125} -\gamma_{25134} + \gamma_{12345} ).
\label{eq:coeffsan4dp}
  \end{align}

  The coefficients of Eq.~\eqref{eq:coeffsan4dp} enter the integrand
  decomposition of Eq.~\eqref{Eq:Multipole4C}.  These results are in
  agreement with the numerical computation of
  Section~\ref{SSec:NumPentatwist}.

\subsection{Higher-rank integrands}
\label{sec:higher-rank}
The analytic reduction via Laurent expansion can be extended to
numerators of higher rank.  As an example we perform the computation
of the sevenfold residues for the $\mathcal{N}=8$ SUGRA two-loop five-points planar pentabox, whose numerator has rank two in the loop
momentum $q$.


The most general parametrization of the residue of the generic sevenfold cut $(i_1\cdots i_7)$ is
\bea
  \Delta_{i_1\cdots i_7}(q,k) = \ci{i_1\cdots i_7}{0} + \ci{i_1\cdots i_7}{1}(q+\mo_0)\cdot \mo_1 + \ci{i_1\cdots i_7}{2} (q+\mo_0)\cdot \mo_2.
 \label{Eq:decR2}
\eea
The momenta $\mo_0^\mu$, $\mo_1^\mu$, and $\mo^\mu_2$ are a linear combination of the external momenta and depend on the cut $(i_1\cdots i_7)$. Their actual 
form is not relevant for this discussion  but can be inferred from Table~\ref{fig:5p:7foldcut:1234568}.    We consider two $t$-dependent solutions of the sevenfold cut:
\bea
(q_i^\mu, k_i^\mu) = \left (v^\mu_{q_i, 1} t +  v^\mu_{q_i, 0}, \;  v^\mu_{k_i}   \right ) \quad \mbox{with} \quad =1,2 \; .
\label{Eq:SolR2}
\eea
The coefficients in Eq.~(\ref{Eq:decR2}) can be computed  evaluating~(\ref{Eq:N8a}) at the solutions~(\ref{Eq:SolR2}) 
\bea
\frac{  \mathcal{N}^{(8,a)}_{1\cdots 8}(q_i, k_i) -\Delta_{12345678}(q_i,k_i) }{  D_{i_8}(q_i,k_i)  }  
&=&    \Delta_{i_1\cdots i_7}(q_i,k_i) \nn
&=& \ci{i_1\cdots i_7}{0} + \ci{i_1\cdots i_7}{1}  (v_{q_i,0} + \mo_0 )  \cdot \mo_1  \nn
&& + \ci{i_1\cdots i_7}{2} (v_{q_i,0} + \mo_0 )   \cdot \mo_2 + \ci{i_1\cdots i_7}{1}(v_{q_i, 1} \cdot \mo_1) t \nn
&& + \ci{i_1\cdots i_7}{2} (v_{q_i, 1}\cdot \mo_2) t, \qquad
\label{Eq:CutR2}
\eea
where $D_{i_8}$ is defined in Eq.~(\ref{Eq:Di8}). The Laurent expansion around $t=\infty$  simplifies the computation. 
Indeed in this limit both 
\bea
 \frac{  \mathcal{N}^{(8,a)}_{1\cdots 8}(q_i, k_i) }{  D_{i_8}(q_i,k_i)  }  \qquad \mbox{and} \qquad \frac{ \Delta_{12345678}(q_i,k_i)  }{  D_{i_8}(q_i,k_i)  } 
\eea
have the same polynomial structure of the residue:
\bea
\left .  \frac{  \mathcal{N}^{(8,a)}_{1\cdots 8}(q_i, k_i) }{  D_{i_8}(q_i,k_i)  }   \right  |_{t\to \infty} &=&  \ni{i_1\cdots i_7}{0}^{[i]} +   \ni{i_1\cdots i_7}{1}^{[i]} t + \mathcal{O}\left ( \frac{1}{t}\right ) \nn
\left . \frac{ \Delta_{12345678}(q_i,k_i)  }{  D_{i_8}(q_i,k_i)  } \right  |_{t\to \infty}  &=& \bi{i_1\cdots i_7}{0}^{[i]}  +  \mathcal{O}\left ( \frac{1}{t}\right ) \; . 
\label{Eq:ExpR2}
 \eea 
The expression  of the coefficients is obtained by plugging the expansions~(\ref{Eq:ExpR2}) in Eq.~(\ref{Eq:CutR2}) and by comparing 
both sides. In particular $\ci{i_1\cdots i_7}{1}$ and $\ci{i_1\cdots i_7}{2}$ are the solution of the 
system 
\begin{equation}
  \left\{
  \begin{array}{r l}
    \ci{i_1\cdots i_7}{1} (v_{q_1}\cdot \mo_1) + \ci{i_1\cdots i_7}{2} (v_{q_1}\cdot \mo_2) & = \ni{i_1\cdots i_7}{1}^{[1]}   \\[2ex]
   \ci{i_1\cdots i_7}{1} (v_{q_2}\cdot \mo_1) + \ci{i_1\cdots i_7}{2} (v_{q_2 }\cdot \mo_2) & = \ni{i_1\cdots i_7}{1}^{[2]}   \\
  \end{array} \right. .
\label{eq:n4sugraansystem}
\end{equation}
The coefficient  $\ci{i_1\cdots i_7}{0}$ is given by 
\bea
\label{eq:anc0sugra}
\ci{i_1\cdots i_7}{0} &=& 
\ni{i_1\cdots i_7}{0}^{[1]} - \bi{i_1\cdots i_7}{0}^{[1]}  -   \ci{i_1\cdots i_7}{1} (v_{q_1,0}+\mo_0 )\cdot \mo_1 - \ci{i_1\cdots i_7}{2} (v_{q_1,0 }+ \mo_0)\cdot \mo_2 \nn
&=& \ni{i_1\cdots i_7}{0}^{[2]} - \bi{i_1\cdots i_7}{0}^{[2]}  -   \ci{i_1\cdots i_7}{1} (v_{q_2,0}+\mo_0 )\cdot \mo_1 - \ci{i_1\cdots i_7}{2} (v_{q_2,0 }+ \mo_0)\cdot \mo_2 ,\qquad
\eea
in terms of the functions
\bea
 \bi{i_1\cdots i_7}{0}^{[i]}  = \frac{ \ci{12345678}{1}  (p_1\cdot v_{q_i,1})  }{ 2 ( v_{q_i,0}+P_{i_8} )\cdot v_{q_i,1}  } \; .
\eea

Eq.~\eqref{eq:anc0sugra} shows that the coefficient $\ci{i_1\cdots
  i_7}{0}$ can be written as the constant term $\ni{i_1\cdots
  i_7}{0}^{[i]}$ of the Laurent expansion of the integrand, corrected
by two kinds of contributions.  The first, $\bi{i_1\cdots
  i_7}{0}^{[i]}$, implements the eightfold-cut subtraction as a
correction at the coefficient level.  The other terms are proportional
to the higher-rank coefficients of the same cut found as solutions of
the system in Eq.~\eqref{eq:n4sugraansystem}.

\section{Conclusions}
We recently proposed a new approach for the reduction of scattering 
amplitudes~\cite{Mastrolia:2012an}, based on multivariate polynomial division.
This technique yields the complete integrand decomposition for arbitrary amplitudes, 
regardless of the number of loops. 
In particular it allows for the determination of (the polynomial form of) the residue at 
any multiparticle cut, 
whose knowledge is a mandatory prerequisite for applying 
the integrand-reduction procedure.
We have also shown how the shape of the residues is uniquely 
determined by the on-shell conditions
and, by using the division modulo Gr\"obner basis, 
we have derived a simple integrand recurrence relation
generating the multiparticle pole decomposition for arbitrary multiloop amplitudes.

In the present paper, we applied the new reduction algorithm to
planar and non planar diagrams appearing in the two-loop five-point 
amplitudes in $\mathcal{N}= 4$ SYM and $\mathcal{N} = 8$ SUGRA (in four dimensions), 
whose numerator functions contain up to rank-two terms in the integration momenta.
We determined all polynomial residues parametrizing 
the cuts of the corresponding topologies and subtopologies.
At the same time, the polynomial form of the residues
defines the integral basis for the amplitude  decomposition.   
For the considered cases, we found that the amplitude can be
decomposed in terms of   independent integrals with eight, seven, 
and six denominators.

Our presented approach is well suited for a seminumerical
implementation.  The mathematical framework it is based on is very
general and provides an effective algorithm for the generalization of
the integrand-reduction method to all orders in perturbation theory.

\section*{Acknowledgments}
We thank Simon Badger, Yang Zhang, and Zhibai Zhang for useful discussions, 
and Ulrich Schubert for valuable comments on the manuscript. 
P.M. and T.P.  are supported by the Alexander von Humboldt Foundation,
in the framework of the Sofja Kovaleskaja Award, endowed by the German
Federal Ministry of Education and Research.  The work of G.O. is
supported in part by the National Science Foundation under Grant
No.~PHY-1068550.  G.O. wishes to acknowledge the support of KITP,
Santa Barbara under National Science Foundation Grant No.~PHY-1125915,
and the kind hospitality of the Max-Planck Institut f\"ur Physik in
Munich during the completion of this project.



\bibliographystyle{JHEP} 
\bibliography{references}

\providecommand{\href}[2]{#2}\begingroup\raggedright\begin{thebibliography}{10}

\bibitem{Bern:1994zx}
Z.~Bern, L.~J. Dixon, D.~C. Dunbar, and D.~A. Kosower, {\it {One-Loop n-Point
  Gauge Theory Amplitudes, Unitarity and Collinear Limits}},  {\em Nucl. Phys.}
  {\bf B425} (1994) 217--260,
  [\href{http://xxx.lanl.gov/abs/hep-ph/9403226}{{\tt hep-ph/9403226}}].

\bibitem{Britto:2004nc}
R.~Britto, F.~Cachazo, and B.~Feng, {\it {Generalized unitarity and one-loop
  amplitudes in N = 4 super-Yang-Mills}},  {\em Nucl. Phys.} {\bf B725} (2005)
  275--305, [\href{http://xxx.lanl.gov/abs/hep-th/0412103}{{\tt
  hep-th/0412103}}].

\bibitem{Cachazo:2004kj}
F.~Cachazo, P.~Svrcek, and E.~Witten, {\it {MHV vertices and tree amplitudes in
  gauge theory}},  {\em JHEP} {\bf 09} (2004) 006,
  [\href{http://xxx.lanl.gov/abs/hep-th/0403047}{{\tt hep-th/0403047}}].

\bibitem{Britto:2004ap}
R.~Britto, F.~Cachazo, and B.~Feng, {\it {New Recursion Relations for Tree
  Amplitudes of Gluons}},  {\em Nucl. Phys.} {\bf B715} (2005) 499--522,
  [\href{http://xxx.lanl.gov/abs/hep-th/0412308}{{\tt hep-th/0412308}}].

\bibitem{Ossola:2006us}
G.~Ossola, C.~G. Papadopoulos, and R.~Pittau, {\it {Reducing full one-loop
  amplitudes to scalar integrals at the integrand level}},  {\em Nucl.Phys.}
  {\bf B763} (2007) 147--169,
  [\href{http://xxx.lanl.gov/abs/hep-ph/0609007}{{\tt hep-ph/0609007}}].

\bibitem{Berger:2008sj}
C.~Berger, Z.~Bern, L.~Dixon, F.~Febres~Cordero, D.~Forde, {\em et.~al.}, {\it
  {An Automated Implementation of On-Shell Methods for One-Loop Amplitudes}},
  {\em Phys.Rev.} {\bf D78} (2008) 036003,
  [\href{http://xxx.lanl.gov/abs/0803.4180}{{\tt arXiv:0803.4180}}].

\bibitem{Giele:2008bc}
W.~Giele and G.~Zanderighi, {\it {On the Numerical Evaluation of One-Loop
  Amplitudes: The Gluonic Case}},  {\em JHEP} {\bf 0806} (2008) 038,
  [\href{http://xxx.lanl.gov/abs/0805.2152}{{\tt arXiv:0805.2152}}].

\bibitem{Badger:2010nx}
S.~Badger, B.~Biedermann, and P.~Uwer, {\it {NGluon: A Package to Calculate
  One-loop Multi-gluon Amplitudes}},  {\em Comput.Phys.Commun.} {\bf 182}
  (2011) 1674--1692, [\href{http://xxx.lanl.gov/abs/1011.2900}{{\tt
  arXiv:1011.2900}}].

\bibitem{Bevilacqua:2011xh}
G.~Bevilacqua, M.~Czakon, M.~Garzelli, A.~van Hameren, A.~Kardos, {\em
  et.~al.}, {\it {HELAC-NLO}},  \href{http://xxx.lanl.gov/abs/1110.1499}{{\tt
  arXiv:1110.1499}}.

\bibitem{Hirschi:2011pa}
V.~Hirschi, R.~Frederix, S.~Frixione, M.~V. Garzelli, F.~Maltoni, {\em
  et.~al.}, {\it {Automation of one-loop QCD corrections}},  {\em JHEP} {\bf
  1105} (2011) 044, [\href{http://xxx.lanl.gov/abs/1103.0621}{{\tt
  arXiv:1103.0621}}].

\bibitem{Cullen:2011ac}
G.~Cullen, N.~Greiner, G.~Heinrich, G.~Luisoni, P.~Mastrolia, {\em et.~al.},
  {\it {Automated One-Loop Calculations with GoSam}},  {\em Eur.Phys.J.} {\bf
  C72} (2012) 1889, [\href{http://xxx.lanl.gov/abs/1111.2034}{{\tt
  arXiv:1111.2034}}].

\bibitem{Agrawal:2011tm}
S.~Agrawal, T.~Hahn, and E.~Mirabella, {\it {FormCalc 7}},
  \href{http://xxx.lanl.gov/abs/1112.0124}{{\tt arXiv:1112.0124}}.

\bibitem{Cascioli:2011va}
F.~Cascioli, P.~Maierhofer, and S.~Pozzorini, {\it {Scattering Amplitudes with
  Open Loops}},  {\em Phys.Rev.Lett.} {\bf 108} (2012) 111601,
  [\href{http://xxx.lanl.gov/abs/1111.5206}{{\tt arXiv:1111.5206}}].

\bibitem{Badger:2012pg}
S.~Badger, B.~Biedermann, P.~Uwer, and V.~Yundin, {\it {Numerical evaluation of
  virtual corrections to multi-jet production in massless QCD}},
  \href{http://xxx.lanl.gov/abs/1209.0100}{{\tt arXiv:1209.0100}}.

\bibitem{Cachazo:2004dr}
F.~Cachazo, {\it {Holomorphic anomaly of unitarity cuts and one-loop gauge
  theory amplitudes}},  \href{http://xxx.lanl.gov/abs/hep-th/0410077}{{\tt
  hep-th/0410077}}.

\bibitem{Britto:2004nj}
R.~Britto, F.~Cachazo, and B.~Feng, {\it {Computing one-loop amplitudes from
  the holomorphic anomaly of unitarity cuts}},  {\em Phys.Rev.} {\bf D71}
  (2005) 025012, [\href{http://xxx.lanl.gov/abs/hep-th/0410179}{{\tt
  hep-th/0410179}}].

\bibitem{Britto:2005ha}
R.~Britto, E.~Buchbinder, F.~Cachazo, and B.~Feng, {\it {One-loop amplitudes of
  gluons in SQCD}},  {\em Phys.Rev.} {\bf D72} (2005) 065012,
  [\href{http://xxx.lanl.gov/abs/hep-ph/0503132}{{\tt hep-ph/0503132}}].

\bibitem{Britto:2006sj}
R.~Britto, B.~Feng, and P.~Mastrolia, {\it {The Cut-constructible part of QCD
  amplitudes}},  {\em Phys.Rev.} {\bf D73} (2006) 105004,
  [\href{http://xxx.lanl.gov/abs/hep-ph/0602178}{{\tt hep-ph/0602178}}].

\bibitem{Forde:2007mi}
D.~Forde, {\it {Direct extraction of one-loop integral coefficients}},  {\em
  Phys. Rev.} {\bf D75} (2007) 125019,
  [\href{http://xxx.lanl.gov/abs/0704.1835}{{\tt arXiv:0704.1835}}].

\bibitem{Badger:2008cm}
S.~D. Badger, {\it {Direct Extraction Of One Loop Rational Terms}},  {\em JHEP}
  {\bf 01} (2009) 049, [\href{http://xxx.lanl.gov/abs/0806.4600}{{\tt
  arXiv:0806.4600}}].

\bibitem{ArkaniHamed:2008gz}
N.~Arkani-Hamed, F.~Cachazo, and J.~Kaplan, {\it {What is the Simplest Quantum
  Field Theory?}},  {\em JHEP} {\bf 1009} (2010) 016,
  [\href{http://xxx.lanl.gov/abs/0808.1446}{{\tt arXiv:0808.1446}}].

\bibitem{Mastrolia:2009dr}
P.~Mastrolia, {\it {Double-Cut of Scattering Amplitudes and Stokes' Theorem}},
  {\em Phys.Lett.} {\bf B678} (2009) 246--249,
  [\href{http://xxx.lanl.gov/abs/0905.2909}{{\tt arXiv:0905.2909}}].

\bibitem{Britto:2010um}
R.~Britto and E.~Mirabella, {\it {Single Cut Integration}},  {\em JHEP} {\bf
  1101} (2011) 135, [\href{http://xxx.lanl.gov/abs/1011.2344}{{\tt
  arXiv:1011.2344}}].

\bibitem{ArkaniHamed:2009dn}
N.~Arkani-Hamed, F.~Cachazo, C.~Cheung, and J.~Kaplan, {\it {A Duality For The
  S Matrix}},  {\em JHEP} {\bf 1003} (2010) 020,
  [\href{http://xxx.lanl.gov/abs/0907.5418}{{\tt arXiv:0907.5418}}].

\bibitem{Ellis:2011cr}
R.~Ellis, Z.~Kunszt, K.~Melnikov, and G.~Zanderighi, {\it {One-loop
  calculations in quantum field theory: from Feynman diagrams to unitarity
  cuts}},  \href{http://xxx.lanl.gov/abs/1105.4319}{{\tt arXiv:1105.4319}}.

\bibitem{Alday:2008yw}
L.~F. Alday and R.~Roiban, {\it {Scattering Amplitudes, Wilson Loops and the
  String/Gauge Theory Correspondence}},  {\em Phys.Rept.} {\bf 468} (2008)
  153--211, [\href{http://xxx.lanl.gov/abs/0807.1889}{{\tt arXiv:0807.1889}}].

\bibitem{Britto:2010xq}
R.~Britto, {\it {Loop Amplitudes in Gauge Theories: Modern Analytic
  Approaches}},  {\em J.Phys.A} {\bf A44} (2011) 454006,
  [\href{http://xxx.lanl.gov/abs/1012.4493}{{\tt arXiv:1012.4493}}]. 34 pages.
  Invited review for a special issue of Journal of Physics A devoted to
  'Scattering Amplitudes in Gauge Theories'.

\bibitem{Henn:2011xk}
J.~M. Henn, {\it {Dual conformal symmetry at loop level: massive
  regularization}},  {\em J.Phys.A} {\bf A44} (2011) 454011,
  [\href{http://xxx.lanl.gov/abs/1103.1016}{{\tt arXiv:1103.1016}}].

\bibitem{Bern:2011qt}
Z.~Bern and Y.-t. Huang, {\it {Basics of Generalized Unitarity}},  {\em
  J.Phys.A} {\bf A44} (2011) 454003,
  [\href{http://xxx.lanl.gov/abs/1103.1869}{{\tt arXiv:1103.1869}}].

\bibitem{Carrasco:2011hw}
J.~J.~M. Carrasco and H.~Johansson, {\it {Generic multiloop methods and
  application to N=4 super-Yang-Mills}},  {\em J.Phys.A} {\bf A44} (2011)
  454004, [\href{http://xxx.lanl.gov/abs/1103.3298}{{\tt arXiv:1103.3298}}].

\bibitem{Dixon:2011xs}
L.~J. Dixon, {\it {Scattering amplitudes: the most perfect microscopic
  structures in the universe}},  {\em J.Phys.A} {\bf A44} (2011) 454001,
  [\href{http://xxx.lanl.gov/abs/1105.0771}{{\tt arXiv:1105.0771}}].

\bibitem{Ita:2011hi}
H.~Ita, {\it {Susy Theories and QCD: Numerical Approaches}},  {\em J.Phys.A}
  {\bf A44} (2011) 454005, [\href{http://xxx.lanl.gov/abs/1109.6527}{{\tt
  arXiv:1109.6527}}].

\bibitem{Bern:1997nh}
Z.~Bern, J.~Rozowsky, and B.~Yan, {\it {Two loop four gluon amplitudes in N=4
  superYang-Mills}},  {\em Phys.Lett.} {\bf B401} (1997) 273--282,
  [\href{http://xxx.lanl.gov/abs/hep-ph/9702424}{{\tt hep-ph/9702424}}].

\bibitem{Bern:2000dn}
Z.~Bern, L.~J. Dixon, and D.~Kosower, {\it {A Two loop four gluon helicity
  amplitude in QCD}},  {\em JHEP} {\bf 0001} (2000) 027,
  [\href{http://xxx.lanl.gov/abs/hep-ph/0001001}{{\tt hep-ph/0001001}}].

\bibitem{Buchbinder:2005wp}
E.~I. Buchbinder and F.~Cachazo, {\it {Two-loop amplitudes of gluons and
  octa-cuts in N=4 super Yang-Mills}},  {\em JHEP} {\bf 0511} (2005) 036,
  [\href{http://xxx.lanl.gov/abs/hep-th/0506126}{{\tt hep-th/0506126}}].

\bibitem{Cachazo:2008vp}
F.~Cachazo, {\it {Sharpening The Leading Singularity}},
  \href{http://xxx.lanl.gov/abs/0803.1988}{{\tt arXiv:0803.1988}}.

\bibitem{Bern:2007ct}
Z.~Bern, J.~Carrasco, H.~Johansson, and D.~Kosower, {\it {Maximally
  supersymmetric planar Yang-Mills amplitudes at five loops}},  {\em Phys.Rev.}
  {\bf D76} (2007) 125020, [\href{http://xxx.lanl.gov/abs/0705.1864}{{\tt
  arXiv:0705.1864}}].

\bibitem{Kosower:2011ty}
D.~A. Kosower and K.~J. Larsen, {\it {Maximal Unitarity at Two Loops}},  {\em
  Phys.Rev.} {\bf D85} (2012) 045017,
  [\href{http://xxx.lanl.gov/abs/1108.1180}{{\tt arXiv:1108.1180}}].

\bibitem{Larsen:2012sx}
K.~J. Larsen, {\it {Global Poles of the Two-Loop Six-Point N=4 SYM integrand}},
   \href{http://xxx.lanl.gov/abs/1205.0297}{{\tt arXiv:1205.0297}}.

\bibitem{CaronHuot:2012ab}
S.~Caron-Huot and K.~J. Larsen, {\it {Uniqueness of two-loop master contours}},
   \href{http://xxx.lanl.gov/abs/1205.0801}{{\tt arXiv:1205.0801}}.

\bibitem{Johansson:2012zv}
H.~Johansson, D.~A. Kosower, and K.~J. Larsen, {\it {Two-Loop Maximal Unitarity
  with External Masses}},  \href{http://xxx.lanl.gov/abs/1208.1754}{{\tt
  arXiv:1208.1754}}.

\bibitem{Mastrolia:2011pr}
P.~Mastrolia and G.~Ossola, {\it {On the Integrand-Reduction Method for
  Two-Loop Scattering Amplitudes}},  {\em JHEP} {\bf 1111} (2011) 014,
  [\href{http://xxx.lanl.gov/abs/1107.6041}{{\tt arXiv:1107.6041}}].

\bibitem{Badger:2012dp}
S.~Badger, H.~Frellesvig, and Y.~Zhang, {\it {Hepta-Cuts of Two-Loop Scattering
  Amplitudes}},  \href{http://xxx.lanl.gov/abs/1202.2019}{{\tt
  arXiv:1202.2019}}.

\bibitem{Zhang:2012ce}
Y.~Zhang, {\it {Integrand-Level Reduction of Loop Amplitudes by Computational
  Algebraic Geometry Methods}},  \href{http://xxx.lanl.gov/abs/1205.5707}{{\tt
  arXiv:1205.5707}}.

\bibitem{Mastrolia:2012an}
P.~Mastrolia, E.~Mirabella, G.~Ossola, and T.~Peraro, {\it {Scattering
  Amplitudes from Multivariate Polynomial Division}},
  \href{http://xxx.lanl.gov/abs/1205.7087}{{\tt arXiv:1205.7087}}.

\bibitem{Tkachov:1981wb}
F.~Tkachov, {\it {A Theorem on Analytical Calculability of Four Loop
  Renormalization Group Functions}},  {\em Phys.Lett.} {\bf B100} (1981)
  65--68.

\bibitem{Gehrmann:2000xj}
T.~Gehrmann and E.~Remiddi, {\it {Using differential equations to compute two
  loop box integrals}},  {\em Nucl.Phys.Proc.Suppl.} {\bf 89} (2000) 251--255,
  [\href{http://xxx.lanl.gov/abs/hep-ph/0005232}{{\tt hep-ph/0005232}}].

\bibitem{Gluza:2010ws}
J.~Gluza, K.~Kajda, and D.~A. Kosower, {\it {Towards a Basis for Planar
  Two-Loop Integrals}},  {\em Phys.Rev.} {\bf D83} (2011) 045012,
  [\href{http://xxx.lanl.gov/abs/1009.0472}{{\tt arXiv:1009.0472}}].

\bibitem{Badger:2012dv}
S.~Badger, H.~Frellesvig, and Y.~Zhang, {\it {An Integrand Reconstruction
  Method for Three-Loop Amplitudes}},  {\em JHEP} {\bf 1208} (2012) 065,
  [\href{http://xxx.lanl.gov/abs/1207.2976}{{\tt arXiv:1207.2976}}].

\bibitem{Kleiss:2012yv}
R.~H. Kleiss, I.~Malamos, C.~G. Papadopoulos, and R.~Verheyen, {\it {Counting
  to one: reducibility of one- and two-loop amplitudes at the integrand
  level}},  \href{http://xxx.lanl.gov/abs/1206.4180}{{\tt arXiv:1206.4180}}.

\bibitem{Feng:2012bm}
B.~Feng and R.~Huang, {\it {The classification of two-loop integrand basis in
  pure four-dimension}},  \href{http://xxx.lanl.gov/abs/1209.3747}{{\tt
  arXiv:1209.3747}}.

\bibitem{Bern:2006ew}
Z.~Bern, M.~Czakon, L.~J. Dixon, D.~A. Kosower, and V.~A. Smirnov, {\it {The
  Four-Loop Planar Amplitude and Cusp Anomalous Dimension in Maximally
  Supersymmetric Yang-Mills Theory}},  {\em Phys.Rev.} {\bf D75} (2007) 085010,
  [\href{http://xxx.lanl.gov/abs/hep-th/0610248}{{\tt hep-th/0610248}}].

\bibitem{Carrasco:2011mn}
J.~J. Carrasco and H.~Johansson, {\it {Five-Point Amplitudes in N=4
  Super-Yang-Mills Theory and N=8 Supergravity}},  {\em Phys.Rev.} {\bf D85}
  (2012) 025006, [\href{http://xxx.lanl.gov/abs/1106.4711}{{\tt
  arXiv:1106.4711}}].

\bibitem{Mastrolia:2012bu}
P.~Mastrolia, E.~Mirabella, and T.~Peraro, {\it {Integrand reduction of
  one-loop scattering amplitudes through Laurent series expansion}},
  \href{http://xxx.lanl.gov/abs/1203.0291}{{\tt arXiv:1203.0291}}.

\bibitem{Vermaseren:2000nd}
J.~A.~M. Vermaseren, {\it {New features of FORM}},
  \href{http://xxx.lanl.gov/abs/math-ph/0010025}{{\tt math-ph/0010025}}.

\bibitem{Maitre:2007jq}
D.~Maitre and P.~Mastrolia, {\it {S@M, a Mathematica Implementation of the
  Spinor-Helicity Formalism}},  {\em Comput. Phys. Commun.} {\bf 179} (2008)
  501--574, [\href{http://xxx.lanl.gov/abs/0710.5559}{{\tt arXiv:0710.5559}}].

\bibitem{Bern:2002tk}
Z.~Bern, A.~De~Freitas, and L.~J. Dixon, {\it {Two loop helicity amplitudes for
  gluon-gluon scattering in QCD and supersymmetric Yang-Mills theory}},  {\em
  JHEP} {\bf 0203} (2002) 018,
  [\href{http://xxx.lanl.gov/abs/hep-ph/0201161}{{\tt hep-ph/0201161}}].

\bibitem{Ossola:2007bb}
G.~Ossola, C.~G. Papadopoulos, and R.~Pittau, {\it {Numerical evaluation of
  six-photon amplitudes}},  {\em JHEP} {\bf 0707} (2007) 085,
  [\href{http://xxx.lanl.gov/abs/0704.1271}{{\tt arXiv:0704.1271}}].

\bibitem{Ossola:2007ax}
G.~Ossola, C.~G. Papadopoulos, and R.~Pittau, {\it {CutTools: a program
  implementing the OPP reduction method to compute one-loop amplitudes}},  {\em
  JHEP} {\bf 03} (2008) 042, [\href{http://xxx.lanl.gov/abs/0711.3596}{{\tt
  arXiv:0711.3596}}].

\bibitem{Ossola:2008xq}
G.~Ossola, C.~G. Papadopoulos, and R.~Pittau, {\it {On the Rational Terms of
  the one-loop amplitudes}},  {\em JHEP} {\bf 0805} (2008) 004,
  [\href{http://xxx.lanl.gov/abs/0802.1876}{{\tt arXiv:0802.1876}}].

\bibitem{Ellis:2007br}
R.~K. Ellis, W.~T. Giele, and Z.~Kunszt, {\it {A Numerical Unitarity Formalism
  for Evaluating One-Loop Amplitudes}},  {\em JHEP} {\bf 03} (2008) 003,
  [\href{http://xxx.lanl.gov/abs/0708.2398}{{\tt arXiv:0708.2398}}].

\bibitem{Giele:2008ve}
W.~T. Giele, Z.~Kunszt, and K.~Melnikov, {\it {Full one-loop amplitudes from
  tree amplitudes}},  {\em JHEP} {\bf 0804} (2008) 049,
  [\href{http://xxx.lanl.gov/abs/0801.2237}{{\tt arXiv:0801.2237}}].

\bibitem{Ellis:2008ir}
R.~Ellis, W.~T. Giele, Z.~Kunszt, and K.~Melnikov, {\it {Masses, fermions and
  generalized $D$-dimensional unitarity}},  {\em Nucl.Phys.} {\bf B822} (2009)
  270--282, [\href{http://xxx.lanl.gov/abs/0806.3467}{{\tt arXiv:0806.3467}}].

\bibitem{Mastrolia:2008jb}
P.~Mastrolia, G.~Ossola, C.~Papadopoulos, and R.~Pittau, {\it {Optimizing the
  Reduction of One-Loop Amplitudes}},  {\em JHEP} {\bf 0806} (2008) 030,
  [\href{http://xxx.lanl.gov/abs/0803.3964}{{\tt arXiv:0803.3964}}].

\bibitem{Mastrolia:2010nb}
P.~Mastrolia, G.~Ossola, T.~Reiter, and F.~Tramontano, {\it {Scattering
  AMplitudes from Unitarity-based Reduction Algorithm at the Integrand-level}},
   {\em JHEP} {\bf 1008} (2010) 080,
  [\href{http://xxx.lanl.gov/abs/1006.0710}{{\tt arXiv:1006.0710}}].

\bibitem{Boughezal:2011br}
R.~Boughezal, K.~Melnikov, and F.~Petriello, {\it {The four-dimensional
  helicity scheme and dimensional reconstruction}},  {\em Phys.Rev.} {\bf D84}
  (2011) 034044, [\href{http://xxx.lanl.gov/abs/1106.5520}{{\tt
  arXiv:1106.5520}}].

\end{thebibliography}\endgroup

\end{document}